arXiv:1608.01077v1 [cond-mat.mtrl-sci] 3 Aug 2016

# Optimal design of low-frequency band gaps in anti-tetrachiral lattice meta-materials

Andrea Bacigalupo[a,*], Giorgio Gnecco[a], Marco Lepidi[b], Luigi Gambarotta[b]

[a] *IMT School for Advanced Studies Lucca, Piazza S. Francesco 19, 55100 Lucca (Italy)*
[b] *DICCA - Dipartimento di Ingegneria Civile, Chimica e Ambientale, Università di Genova, Via Montallegro 1, 16145 Genova (Italy)*

**Abstract**

The elastic wave propagation is investigated in the beam lattice material characterized by a square periodic cell with anti-tetrachiral microstructure. With reference to the Floquet-Bloch spectrum, focus is made on the band structure enrichments and modifications which can be achieved by equipping the cellular microstructure with tunable local resonators. By virtue of its composite mechanical nature, the so-built inertial meta-material gains enhanced capacities of passive frequency-band filtering. Indeed the number, placement and properties of the inertial resonators can be designed to open, shift and enlarge the band gaps between one or more pairs of consecutive branches in the frequency spectrum. In order to improve the meta-material performance, a nonlinear optimization problem is formulated. The maximum of the largest band gap amplitudes in the low-frequency range is selected as suited objective function. Proper inequality constraints are introduced to restrict the optimal solutions within a compact set of mechanical and geometric parameters, including only physically realistic properties of both the lattice and resonators. The optimization problems related to full and partial band gaps are solved independently, by means of a globally convergent version of the numerical method of moving asymptotes, combined with a quasi-Monte Carlo multi-start technique. The optimal solutions are discussed and compared from the qualitative and quantitative viewpoints, bringing to light the limits and potential of the meta-material performance. The clearest trends emerging from the numerical analyses are pointed out and interpreted from the physical viewpoint. Finally, some specific recommendations about the microstructural design of the meta-material are synthesized.

*Corresponding author
*Email address:* andrea.bacigalupo@imtlucca.it (Andrea Bacigalupo)

## 1. Introduction

Composite lattice structures are encountering an increasing success in a number of advanced applications within established and emerging engineering fields. If the mechanical virtues of lattice structures can traditionally be attributed to their optimal material usage and their high designable properties, a growing research attention is currently being focused on their unconventional functional performances. Indeed, lattice materials and meta-materials may offer natural and unique attitudes to develop valuable and adaptable capacities of spatial morphing, remote sensing, health monitoring, active damping, energy harvesting.

Traditional structural realizations of spatial beam lattices include tetrahedral trusses, double-curvature cable networks and hexagonal honeycombs [1–3]. Starting from this well-defined background, a deepening interest has been attracted in the last decades by the geometrical and mechanical design of periodic lattices, with challenging perspectives towards the employment multi-scale hierarchical schemes and the achievement of multiphysical functionalities [4–7]. As major findings of this optimization trend, novel promising architectural *topologies* have been discovered and new challenging structural *typologies* have been proposed. Among the novel geometric topologies – for instance – the cellular layouts based on chiral and anti-chiral cellular symmetries have demonstrated distinctive elastic properties, such as a marked auxeticity and shear rigidity, together with outstanding capacities of fracture toughness, indentation resistance and energy absorption [8–16]. Among the new structural typologies – for instance – the tri-dimensional tendon-strut systems based on the tensegrity concept have virtuously conjugated remarkable strength-to-lightness ratios with extreme properties of softening/hardening elasticity and high degrees of spatial deployability [17–20].

Chiral cellular materials have also drawn many research efforts specifically focused on their high design flexibility as fully mechanical filters for the transmission and dispersion of elastic waves [21–23]. Indeed, the chiral microstructures, characterized by a composite pattern of rings and ligaments, are suited for fine-grained customizations of the acoustic material properties. Employing the Floquet-Bloch theory for periodic media [24–27], the band structure properties of different chiral materials have been extensively studied, with reference to beam lattice microstructural models [21, 22, 28] as well as to equivalent homogenized continua [23, 29]. Within this framework, enhanced capabilities of frequency band filtering can be achieved by the *inertial meta-materials*, which leverage the negative effective mass density that can be obtained with the introduction of intra-ring massive disks [30–32]. By virtue of the ring-disk elastic coupling, these auxiliary oscillators works as local inertial resonators which – if properly tuned – can open, shift and enlarge the spectrum band gaps in response to specific design requirements [33–35].

In the rich library of chiral topologies, the class of anti-tetrachiral materials, first identified as the structural solution of a topological optimization problem [36], is highly attractive for its strong auxeticity, accompanied by a marked anisotropy of the elasto-dynamic response [9, 11, 29, 37–39]. In terms of dynamic analyses, the micro-structural complexity of the anti-tetrachiral periodic cell requires several independent parameters for a minimal but complete description of the elasto-dynamic material behavior. Furthermore, the tetra-atomic cellular configuration implies a quadruple number of Lagrangian coordinates with respect to the mono-atomic layout of the trichiral, tetrachiral or hexachiral materials. Consequently, the band structure of the material possess a high spectral density, with several dispersion curves interacting to each other in the same frequency range, and a strong sensitivity of the spectral properties to a number of structural parameters. To date, the wave propagation spectrum of the anti-tetrachiral material has been determined by solving the eigenproblem governing the Floquet-Bloch theory through numerical continuation methods or asymptotic perturbation techniques [28, 29]. Some preliminary parametric analyses have also confirmed the possibility to control the band structures by means of tuned inertial resonators [40]. Nonetheless, the high spectral density and the large variety of design parameters make the anti-tetrachiral meta-material a challenging benchmark for the application of structural optimization strategies oriented to improve and likely maximize its performance as mechanical filter.

Based on these motivations, the present paper focuses on the spectral optimization of the anti-tetrachiral periodic meta-material, with the objective to maximize the amplitude of the low-frequency band gap in the Floquet-Bloch spectrum. Fixed the anti-tetrachiral topology, the optimization concerns the key parameters describing both the microstructural model of the periodic cell and the elasto-dynamic properties of the inertial resonators. The number and placement of the resonators are accounted for as complementary optimization variables. The paper is organized as follows. The physical-mathematical description of the periodic meta-material is formulated in the framework of linear elasticity (Section 2). In particular, a beam lattice model is defined to govern the free undamped vibration of the periodic cell (Section 2). Therefore, the Floquet-Bloch theory is invoked to describe the wave propagation, and the significant spectral properties are expressed as function of a physically admissible set of mechanical parameters (Section 2.2). The optimization problem is outlined (Section 3), and then formally stated with proper distinction between the maximization of full and partial band gaps (Section 3.1). After a brief introduction to the numerical solution methods, which are featured by global convergence and quasi-random initialization (Section 3.2), two scenarios or levels of optimization problems are tackled. In the first scenario, all the possible cell configurations, which differ to each other for the number and placement of the resonators, are independently analyzed and optimized (Section 3.3). In the second scenario, the best cell configuration is fixed and a further optimization process is focused only on the resonator properties (Section 3.4). The optimization results are discussed from a qualitative and quantitative viewpoint, some design recommendations are pointed out on the base of the clearest trends emerged from the analyses and, finally, concluding remarks are offered.

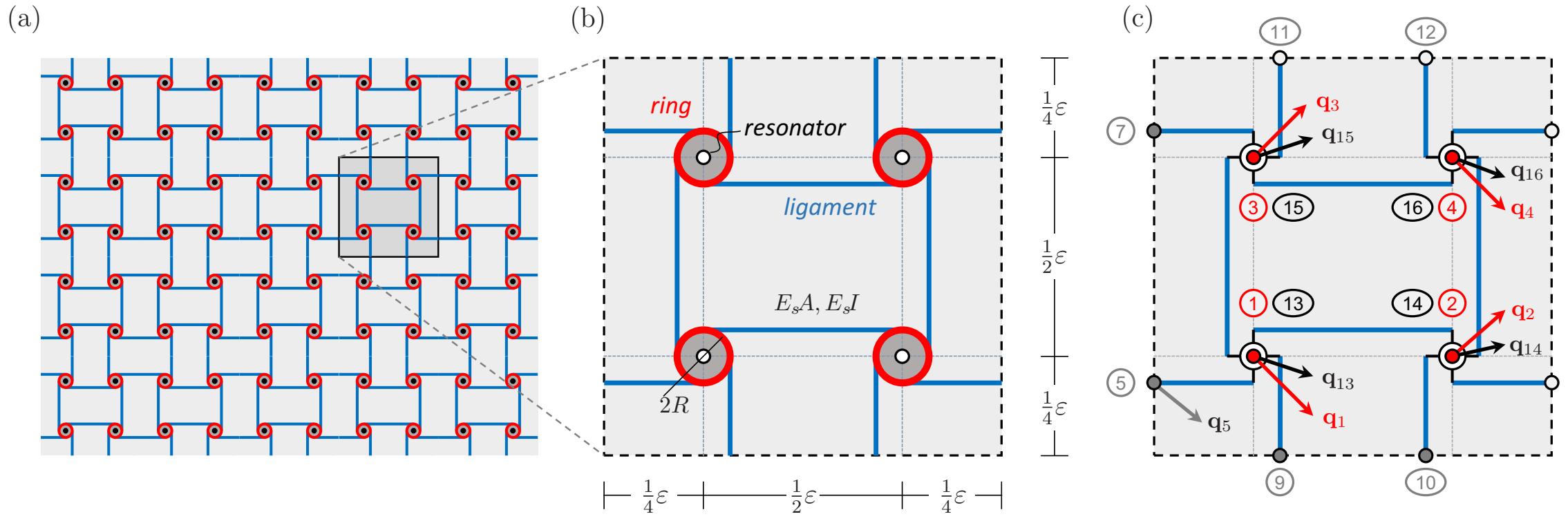

Figure 1: Anti-tetrachiral cellular meta-material: (a) pattern, (b) periodic cell, (c) beam lattice model.

## 2. Anti-tetrachiral material vs meta-material

A composite periodic material is characterized by square cells, that realize the regular tiling of an infinite bi-dimensional domain (Figure 1a). The geometric and mechanical properties of the generic cell are featured by a double symmetry, according to an anti-tetrachiral planar topology (Figure 1b). The cell microstructure, with characteristic size $\varepsilon$, is composed by four circular rings connected by twelve tangent ligaments. The *rolling-up* mechanism, responsible for the auxetic behavior, consists in the opposite-sign, iso-amplitude rotations developed by any pair of adjacent rings in-a-row (or column), when the cell is stretched along one of the symmetry axes. With respect to the traditional anti-tetrachiral material, the introduction of intra-ring (mass-in-mass) resonators can give birth to a high-performing inertial meta-material, characterized by enhanced capacities of filtering low-frequency bands of elastic harmonic waves.

### *2.1. Beam lattice model*

A beam lattice model is formulated to describe the linear elasto-dynamic response of the cellular composite with unit thickness. A rigid body model is assumed for the massive and highly-stiff rings, possessing identical mean radius $R$. The ring centers are located at the four corners of an ideal internal square, concentric with the external cell boundary. The ring width $W$ is considered a free parameter, allowing the independent assignment of the rigid body mass $M$ and moment of inertia $J$, fixed the material density $\varrho$. A linear, extensible, unshearable model of massless beam is employed for all the light and flexible ligaments, in the small-deformation range. As long as the beam-ring connections nominally realize perfectly-rigid joints, the natural length $L$ of the *inner* horizontal and vertical ligaments coincides with half the side of the square cell. By virtue of the periodicity, the cell boundary crosses the midspan – and halves the natural length – of all the *outer* ligaments. Assuming a linear elastic material (with Young's modulus $E_s$) and a rectangular cross-section shape (with area $A$ and second area moment $I$ depending on the width $w_s$) for each ligament, all the beams have identical extensional $E_sA$ and flexural rigidity $E_sI$. The effects of a homogeneous soft matrix, which may likely embed the microstructure [29], are neglected as first approximation.

Moving from this structural layout, a meta-material can be realized by supplying each ring with a light soft annular filler, hosting a central heavy circular inclusion, serving as inertial resonator with adjustable mechanical properties. All the inclusions are assumed identical and highly stiff, so that they can be modelled as rigid disks, co-centered with the respective housing rings, with body mass $M_r$ and moment of inertia $J_r$ depending on the radius $R_r$ and the material density $\varrho_r$. As long as the internal coupling provided by the filler can be assumed linearly elastic (with Young's modulus $E_r$ and Poisson ratio $\nu_r$), the ring-resonator differential displacements are affected by equivalent translational and rotational stiffnesses [34].

Employing the direct stiffness method to formulate a parametric Lagrangian model, the free undamped dynamics of the generic periodic cell is governed by a linear system of ordinary differential equations, defined in the full configuration space, which includes three planar degrees-of-freedom (two translations and one rotation) for each of the 16 nodes in the beam lattice model (Figure 1c). Eight external nodes (labeled ⑤...⑫) are located at the midspan of the outer ligaments, crossed by the cell boundary. The remaining eight internal nodes are placed in the centroid positions of the rings (nodes ①...④) and the resonators (nodes ⑬...⑯), which coincide to each other in the natural configuration. Therefore, the model dimension has been properly reduced by virtue of a quasistatic condensation procedure applied to the eight *passive* (mass-free) external nodes lying on the cell boundary. Furthermore, the Floquet-Bloch boundary conditions have been imposed on the condensed model, in order to account for the planar wave propagation across adjacent cells. Finally, the free undamped dynamics of the material is governed by a linear equation defined in the reduced configuration space of the only 24 *active* (massive) internal degrees-of-freedom [28].

The nondimensional configuration vector $\mathbf{q}_a = (\mathbf{q}_s, \mathbf{q}_r)$, joining columnwise the active nodal displacement vectors of the ring $\mathbf{q}_s = (\mathbf{q}_1, ..., \mathbf{q}_4)$ and of the resonator $\mathbf{q}_r = (\mathbf{q}_{13}, ..., \mathbf{q}_{16})$, can be introduced. The displacement vector of the $i$-th active node $(i = 1, ..., 4, 13, ..., 16)$ is

$$\mathbf{q}_i = \left( \frac{u_1^i}{\varepsilon}, \frac{u_2^i}{\varepsilon}, \phi^i \right) \tag{1}$$

where $(u_1^i, u_2^i, \phi^i)$ are the three planar degrees-of-freedom (translations $u_1^i, u_2^i$ and rotation $\phi^i$).

Introducing a convenient reference frequency $\Omega_s$, the nondimensional time $\tau = \Omega_s t$ and the nondimensional wavevector $\mathbf{k} = \varepsilon\,(k_1, k_2)$, the nondimensional form of the equation of motion reads

$$\mathbf{M}(\boldsymbol{\mu})\ddot{\mathbf{q}}_a + \mathbf{K}(\boldsymbol{\mu}, \mathbf{k})\mathbf{q}_a = \mathbf{0} \tag{2}$$

where the dot indicates differentiation with respect to the nondimensional time. More details about the equation formulation are reported in the Appendix A. The hermitian matrices $\mathbf{M}(\boldsymbol{\mu})$ and $\mathbf{K}(\boldsymbol{\mu}, \mathbf{k})$ are related to the mass and stiffness of the periodic cell, and depend on the nondimensional vector of seven independent parameters

$$\boldsymbol{\mu} = \left( \frac{w_s}{\varepsilon}, \frac{W}{w_s}, \frac{R}{\varepsilon}, \frac{R_r}{\varepsilon}, \frac{E_r}{E_s}, \nu_r, \frac{\varrho_r}{\varrho} \right) \tag{3}$$

which belongs to $\mathbb{R}^7$. It is worth noting that the parameters $E_r/E_s$ and $\varrho_r/\varrho$ represent the resonator-to-ligament *elastic ratio* and the resonator-to-ring *mass density ratio*, whereas all the other parameters – except the Poisson ratio $\nu_r$ – account for geometric properties of the periodic cell. The parametric form of the matrices is reported in the Appendix B. In the absence of resonators, the matrices depend on the parameter subvector $\boldsymbol{\mu}_s$, which belongs to $\mathbb{R}^3$ and includes the first three elements of the full vector $\boldsymbol{\mu}$.

### 2.2. Wave propagation

Denoting by $\Omega$ and $\omega = \Omega/\Omega_s$ the unknown dimensional and nondimensional frequencies, the oscillatory solution $\mathbf{q}_a = \boldsymbol{\psi} \exp(\iota\omega\tau)$ can be imposed in the equation of motion. Therefore, eliminating the dependence on time, the in-plane wave propagation is governed by the linear eigenproblem

$$\left(\mathbf{K}(\boldsymbol{\mu},\mathbf{k}) - \omega^2\mathbf{M}(\boldsymbol{\mu})\right)\boldsymbol{\psi} = \mathbf{0} \tag{4}$$

in the unknown eigenvalues $\omega^2$ and eigenvectors $\boldsymbol{\psi}$. The eigenproblem solution is composed by 24 eigenpairs, each made of a real-valued eigenvalue $\omega_h^2(\boldsymbol{\mu},\mathbf{k})$, and a complex-valued eigenvector $\boldsymbol{\psi}_h(\boldsymbol{\mu},\mathbf{k}) \in \mathbb{C}^{24}$ (with $h = 1, ..., 24$). Here, the eigenvector $\boldsymbol{\psi}_h$ is the polarization mode of the $\omega_h$-monofrequent propagating wave. Each eigenvalue can be determined as one of the roots of the characteristic equation

$$\mathcal{P}(\omega,\boldsymbol{\mu},\mathbf{k}) = \det\left(\mathbf{K}(\boldsymbol{\mu},\mathbf{k}) - \omega^2\mathbf{M}(\boldsymbol{\mu})\right) = 0 \tag{5}$$

In the specific case of inertial lattice meta-materials and with reference to the partitioned configuration vector $\mathbf{q}_a = (\mathbf{q}_s, \mathbf{q}_r)$, the following general partition can be applied to the governing matrices

$$\mathbf{K}(\boldsymbol{\mu},\mathbf{k}) = \begin{bmatrix} \mathbf{K}_s(\boldsymbol{\mu},\mathbf{k}) & -\mathbf{K}_r(\boldsymbol{\mu}) \\ -\mathbf{K}_r(\boldsymbol{\mu}) & \mathbf{K}_r(\boldsymbol{\mu}) \end{bmatrix}, \quad \mathbf{M}(\boldsymbol{\mu}) = \begin{bmatrix} \mathbf{M}_s(\boldsymbol{\mu}) & \mathbf{O} \\ \mathbf{O} & \mathbf{M}_r(\boldsymbol{\mu}) \end{bmatrix} \tag{6}$$

where all the sub-matrices are reported in the Appendix B. In particular, the sub-matrices $\mathbf{K}^r(\boldsymbol{\mu})$ and $\mathbf{M}^r(\boldsymbol{\mu})$ are the nondimensional stiffness and mass matrices of the free-standing resonators. For the anti-tetrachiral meta-materials they can be expressed in the diagonal block form

$$\mathbf{K}_r(\boldsymbol{\mu}) = \text{diag}\left[\delta_{13}\mathbf{K}^r_{13}(\boldsymbol{\mu}), \delta_{14}\mathbf{K}^r_{14}(\boldsymbol{\mu}), \delta_{15}\mathbf{K}^r_{15}(\boldsymbol{\mu}), \delta_{16}\mathbf{K}^r_{16}(\boldsymbol{\mu})\right] \tag{7}$$

$$\mathbf{M}_r(\boldsymbol{\mu}) = \text{diag}\left[\mathbf{M}^r_{13}(\boldsymbol{\mu}), \mathbf{M}^r_{14}(\boldsymbol{\mu}), \mathbf{M}^r_{15}(\boldsymbol{\mu}), \mathbf{M}^r_{16}(\boldsymbol{\mu})\right] \tag{8}$$

where $\delta_j = 1$ indicates the presence of the resonator in the $j$-th node (with $j = 13, ..., 16$), whereas its absence is conventionally simulated by artfully zeroing the elastic coupling between the disk and the ring (by setting $\delta_j = 0$). Accordingly, the characteristic equation can be factorized in the form

$$\mathcal{P}(\omega,\boldsymbol{\mu},\mathbf{k}) = \omega^{3(4-n)}\mathcal{Q}_n(\omega,\boldsymbol{\mu},\mathbf{k}) = 0 \tag{9}$$

where $n$ stands for the variable number of resonators in the periodic cell ($n = 1, ..., 4$). This general formulation includes also the absence of resonators as particular case (for $n = 0$).

The dispersion relations $\omega_h(\boldsymbol{\mu},\mathbf{k})$ can be determined as the positive solutions of the equation $\mathcal{Q}_n(\omega,\boldsymbol{\mu},\mathbf{k}) = 0$ for $h = 1, ..., (12 + 3n)$. For any fixed choice of the parameter vector $\boldsymbol{\mu}$, the $h$-th *dispersion surface* of the Floquet-Bloch spectrum is defined as the $h$-th frequency locus $\omega_h(\boldsymbol{\mu},\mathbf{k})$ under variation of the wavevector $\mathbf{k}$ on the whole first Brillouin zone $\mathcal{B} = [-\pi,\pi] \times [-\pi,\pi]$ [24, 34]. Similarly, the $h$-th *dispersion curve* of the Floquet-Bloch spectrum is defined as the $h$-th frequency locus $\omega_h\left(\boldsymbol{\mu},\mathbf{k}\left(\Xi\right)\right)$ along the closed polygonal curve $\Gamma$, defined in $\mathcal{B}$ by the ordered vertices $\mathbf{k}_0=(0,0)$, $\mathbf{k}_1=(\pi,0)$, $\mathbf{k}_2=(\pi,\pi)$, $\mathbf{k}_3=(0,0)$, $\mathbf{k}_4=(0,\pi)$, $\mathbf{k}_5=(\pi,\pi)$, $\mathbf{k}_6=(0,0)$ and spanned by the curvilinear coordinate $\Xi$ (Figure 2b). Therefore, the six $\Gamma$-segments are $\Gamma_1$, $\Gamma_2$, $\Gamma_3$, $\Gamma_4$, $\Gamma_5$, $\Gamma_6$, and correspond to the $\Xi$-intervals $[\Xi_0,\Xi_1]$, $[\Xi_1,\Xi_2]$, $[\Xi_2,\Xi_3]$, $[\Xi_3,\Xi_4]$, $[\Xi_4,\Xi_5]$, $[\Xi_5,\Xi_6]$, where $\Xi_0=0$, $\Xi_1=\pi$, $\Xi_2=2\pi$, $\Xi_3=\left(2+\sqrt{2}\right)\pi$, $\Xi_4=\left(3+\sqrt{2}\right)\pi$, $\Xi_5=\left(4+\sqrt{2}\right)\pi$, $\Xi_6=\left(4+2\sqrt{2}\right)\pi$.

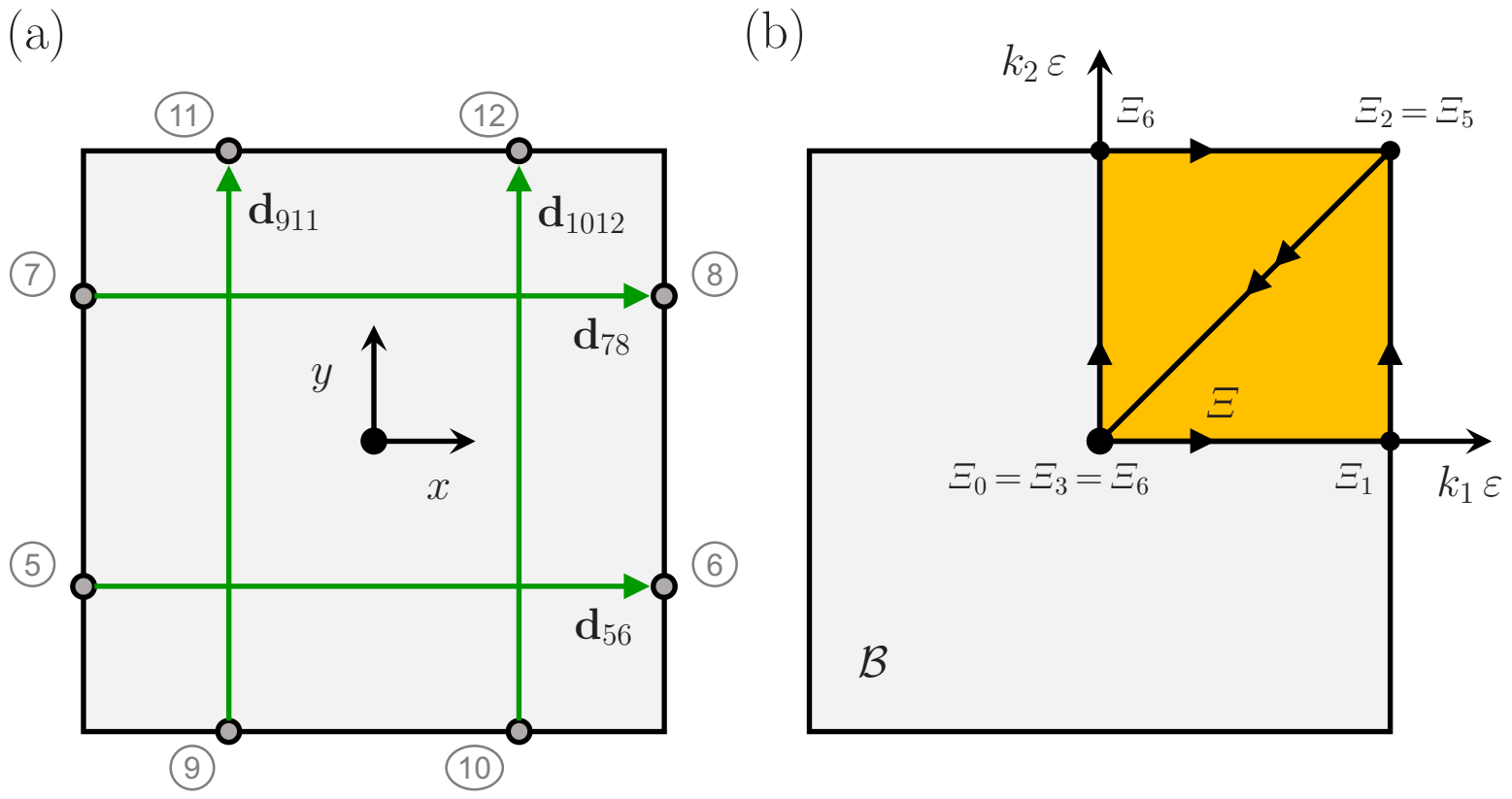

Figure 2: Periodic square cell: (a) central reference system and periodicity vectors, (b) whole first Brillouin zone $\mathcal{B}$ and $\Xi$-coordinate spanning the closed polygonal curve $\Gamma$.

In the Floquet-Bloch spectrum, the frequency range covered by a single dispersion surface is denoted as *frequency band*. Within the $\mathcal{B}$-domain, two dispersion surfaces can cross each other, or – in alternative – be separated by a frequency range referred to as *full band gap*, whereas *partial band gaps* are restricted to a certain $\mathbf{k}$-direction of the $\mathcal{B}$-plane. Sorting the frequencies in ascending order, the relative amplitude of the full band gap (*full amplitude*) between the consecutive $h$-th and $k$-th dispersion surfaces (with $k=h+1$) is

$$\Delta\omega_{hk,\mathcal{B}}\left(\boldsymbol{\mu}\right)=\frac{\max\limits_{\mathbf{k}\in\mathcal{B}}\omega_h(\boldsymbol{\mu},\mathbf{k})-\min\limits_{\mathbf{k}\in\mathcal{B}}\omega_k(\boldsymbol{\mu},\mathbf{k})}{\frac{1}{2}\left[\max\limits_{\mathbf{k}\in\mathcal{B}}\omega_h(\boldsymbol{\mu},\mathbf{k})+\min\limits_{\mathbf{k}\in\mathcal{B}}\omega_k(\boldsymbol{\mu},\mathbf{k})\right]} \tag{10}$$

where the denominator normalizes the amplitude with respect to the mean frequency of the band gap. It is worth noting that the function $\Delta\omega_{hk,\mathcal{B}}\left(\boldsymbol{\mu}\right)$ can attain non-positive values, if the maximum of the $h$-th surface is larger that the minimum of the $k$-th surface. This occurrence corresponds to the absence of full band gap. Similarly, the relative amplitude of the partial band gap (*partial amplitude*) between the consecutive $h$-th and $k$-th dispersion curves, restricted on the segments $\Gamma_p$ (for $p=1,...,6$), takes the form

$$\Delta\omega_{hk,\Gamma_p}\left(\boldsymbol{\mu}\right)=\frac{\max\limits_{\mathbf{k}\in\Gamma_p}\omega_h(\boldsymbol{\mu},\mathbf{k})-\min\limits_{\mathbf{k}\in\Gamma_p}\omega_k(\boldsymbol{\mu},\mathbf{k})}{\frac{1}{2}\left[\max\limits_{\mathbf{k}\in\Gamma_p}\omega_h(\boldsymbol{\mu},\mathbf{k})+\min\limits_{\mathbf{k}\in\Gamma_p}\omega_k(\boldsymbol{\mu},\mathbf{k})\right]} \tag{11}$$

and is associated to waves with wavevector $\mathbf{k}\in\Gamma_p$. Again, the absence of partial band gaps is revealed by non-positive values of the function $\Delta\omega_{hk,\Gamma_p}\left(\boldsymbol{\mu}\right)$.

The variable number of the inertial resonators determine how many non-zero dispersion surfaces constitute the meta-material spectrum, consistently with the solutions of the characteristic equation (9). Therefore, it may be convenient to apply the superscript $(n)$ to the band gap amplitudes $\Delta\omega_{hk}^{(n)}$ (with $n=0,...,4$) to specify the actual number of different evaluable band gaps, which is limited to $2 \leq k \leq (12+3n)$.

Table 1: Lower and upper bounds on the geometrical and mechanical parameters

| $\boldsymbol{\mu}$ | $\frac{w_s}{\varepsilon}$ | $\frac{W}{w_s}$ | $\frac{R}{\varepsilon}$ | $\frac{R_r}{\varepsilon}$ | $\frac{E_r}{E_s}$ | $\nu_r$ | $\frac{\rho_r}{\rho}$ |
|---|---|---|---|---|---|---|---|
| $\boldsymbol{\mu}_{\min}$ | $\frac{3}{100}$ | $\frac{1}{20}$ | $\frac{1}{20}$ | $\frac{1}{100}$ | $\frac{1}{10}$ | $\frac{1}{5}$ | $\frac{1}{2}$ |
| $\boldsymbol{\mu}_{\max}$ | $\frac{1}{10}$ | $\frac{10}{3}$ | $\frac{1}{10}$ | $\frac{9}{100}$ | 1 | $\frac{2}{5}$ | 2 |

The admissible bounds for the parameter vector $\boldsymbol{\mu}$ are the vectors $\boldsymbol{\mu}_{\min}$ and $\boldsymbol{\mu}_{\max}$, whose components are reported in Table 1. The condition $\boldsymbol{\mu}_{\min} \leq \boldsymbol{\mu} \leq \boldsymbol{\mu}_{\max}$ (where the inequalities are interpreted componentwise) defines the bounded parameter space $\mathcal{M} \subset \mathbb{R}^7$ in the presence of resonators. Similarly, the first three components of $\boldsymbol{\mu}_{\min}$ and $\boldsymbol{\mu}_{\max}$ bound the reduced parameter space $\mathcal{M}_s \subset \mathbb{R}^3$ in the absence of resonators. Furthermore, the geometric elements of the parameter vector $\boldsymbol{\mu}$ must obey to some physical restrictions (*geometric constraints*), which can be expressed by the coupled nonlinear inequalities

$$\frac{1}{10}\frac{R}{\varepsilon} \leq \frac{W}{w_s}\frac{w_s}{\varepsilon} \leq \frac{R}{\varepsilon} \tag{12}$$

$$\frac{w_s}{\varepsilon} \leq \frac{2}{3}\left(\frac{R}{\varepsilon} + \frac{1}{2}\frac{W}{w_s}\frac{w_s}{\varepsilon}\right) \tag{13}$$

$$\frac{1}{5}\left(\frac{R}{\varepsilon} - \frac{1}{2}\frac{W}{w_s}\frac{w_s}{\varepsilon}\right) \leq \frac{R_r}{\varepsilon} \leq \frac{9}{10}\left(\frac{R}{\varepsilon} - \frac{1}{2}\frac{W}{w_s}\frac{w_s}{\varepsilon}\right) \tag{14}$$

In the absence of resonators, the constraint (14) is absent. Joining the condition $\boldsymbol{\mu}_{\min} \leq \boldsymbol{\mu} \leq \boldsymbol{\mu}_{\max}$ with these geometric constraints, the *admissible* regions $\mathcal{R} \subseteq \mathcal{M}$ and $\mathcal{R}_s \subseteq \mathcal{M}_s$ of the parameter vectors $\boldsymbol{\mu}$ and $\boldsymbol{\mu}_s$ are respectively defined.

## 3. Band gap optimization

According to a classic approach to inverse problems, a suited objective function $\mathcal{F}(\boldsymbol{\mu})$ has to be defined, and the particular parameter combination $\boldsymbol{\mu}^\circ$ corresponding to its optimal value $\mathcal{F}^\circ = \mathcal{F}(\boldsymbol{\mu}^\circ)$ is sought for. Due to the inherent nonlinearity of the optimization problem, the solution, if it exists in the admissible parameter space $\mathcal{R}$, can be not unique. Furthermore, local and global optima may co-exist. To achieve the highest material performance according to the functional criteria of the present study, the largest amplitude of all the band gaps is selected as objective function. The optimization problem requires its maximization.

For the sake of clarity, it is worth remarking that the above issue differs from other band gap maximization problems, which specifically deal with the topological optimization of phononic materials. Indeed, although pursuing the same objective (the largest band gap amplitude), the *topological optimization* seeks for the optimal distribution of two or more material phases in a sufficiently-fine pixelation of the periodic cell [41–44]. On the contrary, here the anti-tetrachiral topology of the periodic cell is fixed a priori, whereas the *parametric optimization* is limited to the cellular micro-structural parameters, whose values allow to distinguish among different materials belonging to the same topological class. To same extent, the present analyses are aligned with the search for the maximum band gap achievable by varying the connection number and the joint rigidity in periodic lattices made of beam frameworks, in the absence of resonators [45]. It could be also remarked that the optimization problems related to inertial meta-materials equipped with local resonators may have some conceptual and formal similarities with the optimal design of multiple tuned mass dampers for the vibration mitigation in civil and mechanical engineering structures [46–48]

First, the material optimization is performed in the absence of inertial resonators (Case $\mathcal{C}_0$ in Figure 3), in order to explore the potential optimality of the anti-tetrachiral microstructure without additional mechanical resources. Then, the optimization problem is tackled for five different configurations of the meta-material, which differ in the number and placement of the resonators (Cases $\mathcal{C}_1$, $\mathcal{C}_{2h}$, $\mathcal{C}_{2d}$, $\mathcal{C}_3$, $\mathcal{C}_4$ in Figures 3). Indeed, the number and locations of the inertial resonators strongly affects the spectral properties, so that these variables must be considered complementary objects of the optimization problems. It is worth noting that the two-resonator case is distinguished in two sub-cases, depending on the resonator position. Therefore, the superscripts $n = 2h$ and $n = 2d$ refer to the resonators housed by two rings on a row or on a diagonal, respectively. Any other configuration which turns out from permutations of the resonator locations, is spectrally equivalent to one of the five Cases $\mathcal{C}_{1-4}$, by virtue of the structural symmetry and the model linearity. For the sake of consistency, the different meta material configurations are optimized with focus on a fixed number of dispersion surfaces, lying in the low-frequency range. In particular, the lowest twelve frequencies are retained in the band gap amplitude ($h=1,\dots,11$ and $k=2,\dots,12$) used as objective function, so that the qualitative and quantitative results from different cases can be compared to each other.

### *3.1. Optimization problems*

Considering that the first two dispersion surfaces certainly cross the origin of the $(\mathcal{B},\omega)$-space for the selected parameter regions, the related band gap amplitude (for $h=1, k=2$) is always non-positive. Therefore, the search for full band gaps can start from the next pair of dispersion surfaces ($h=2, k=3$).

Focusing first on the full band gaps for the case $\mathcal{C}_0$, the maximization of the full amplitude is governed by the optimization problem

$$
\begin{aligned}
&\underset{\boldsymbol{\mu}_s}{\text{maximize}} && \mathcal{F}_{\mathcal{B}}^{(0)}(\boldsymbol{\mu}_s) \doteq \max_{h=2,\dots,11} \Delta\omega^{(0)}_{h(h+1),\mathcal{B}}(\boldsymbol{\mu}_s) \\
&\text{s.t.} && \boldsymbol{\mu}_{s,\min} \leq \boldsymbol{\mu}_s \leq \boldsymbol{\mu}_{s,\max} \\
& && \text{and the constraints (12) and (13)}
\end{aligned}
\tag{15}
$$

For each admissible $\boldsymbol{\mu}_s$-value, the objective function $\mathcal{F}_{\mathcal{B}}^{(0)}$ expresses the largest full amplitude within all the ten pairs of consecutive dispersion surfaces.

If the problem solution does not discover any full band gap, partial band gaps are searched along selected

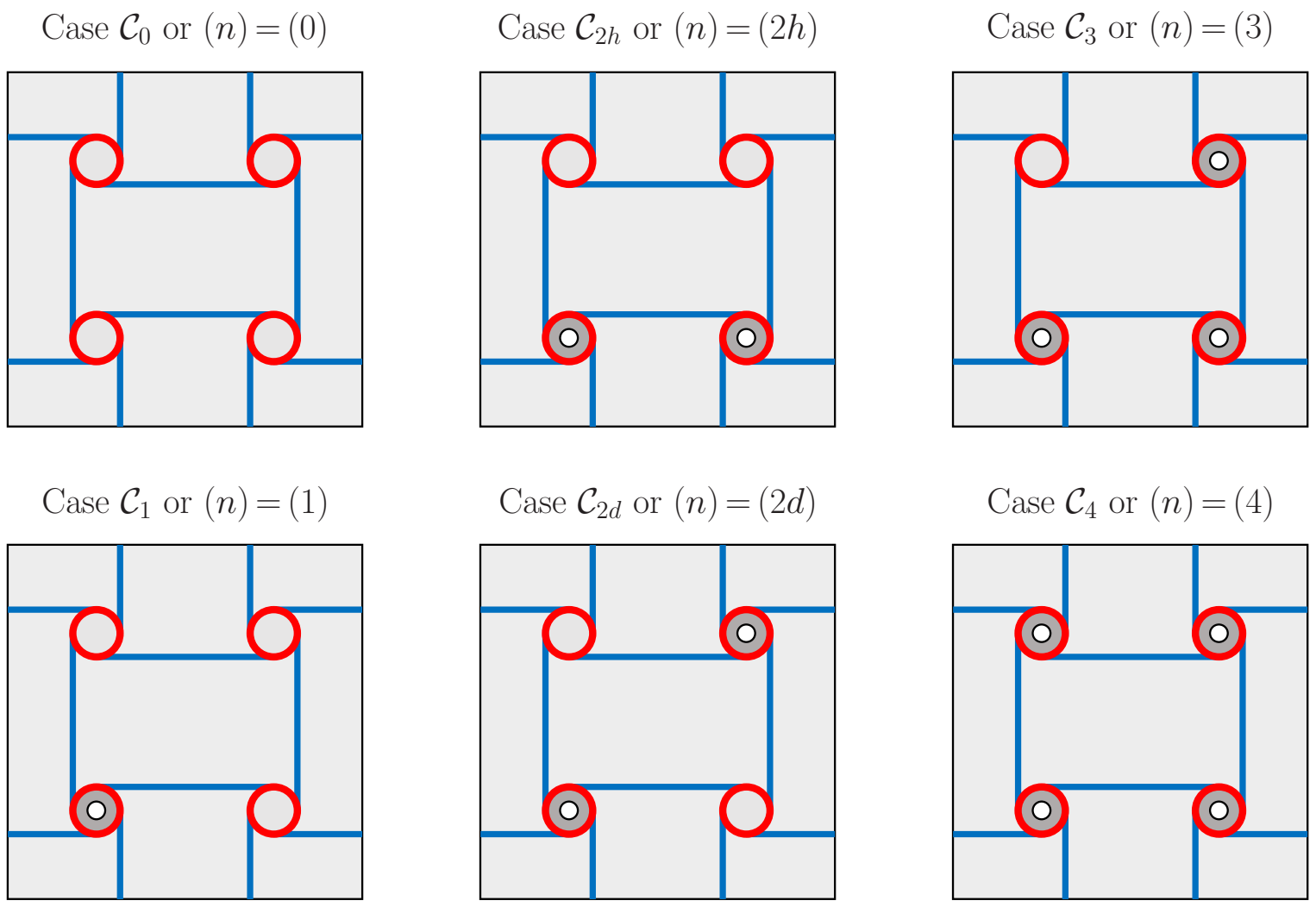

Figure 3: Different configurations of the inertial resonators in the periodic square cell of the meta-material.

segments of the polygonal curve $\Gamma$, by formulating the new optimization problem

$$\begin{aligned} &\underset{\boldsymbol{\mu}_s}{\text{maximize}} && \mathcal{F}_{\Gamma}^{(0)}(\boldsymbol{\mu}_s) \doteq \max_{h=2,\dots,11} \max_{p=1,3,4,6} \Delta\omega^{(0)}_{h(h+1),\Gamma_p}(\boldsymbol{\mu}_s) \\ &\text{s.t.} && \boldsymbol{\mu}_{s,\min} \leq \boldsymbol{\mu}_s \leq \boldsymbol{\mu}_{s,\max} \\ & && \text{and the constraints (12) and (13)} \end{aligned} \tag{16}$$

where the objective function $\mathcal{F}_{\Gamma}^{(0)}$ implies a double maximization with respect to each pair of dispersion curves and each $\Gamma$-segment. The problem solution can – in principle – reveal the absence of partial band gaps. If this is the case, it is concluded that neither full nor partial band gaps exist.

With focus on the band gaps of the remaining cases ($\mathcal{C}_1$, $\mathcal{C}_{2h}$, $\mathcal{C}_{2d}$, $\mathcal{C}_3$, $\mathcal{C}_4$), the maximization of the largest full amplitude is governed by the optimization problem

$$\begin{aligned} &\underset{\boldsymbol{\mu}}{\text{maximize}} && \mathcal{F}_{\mathcal{B}}^{(n)}(\boldsymbol{\mu}) \doteq \max_{h=2,\dots,11} \Delta\omega^{(n)}_{h(h+1),\mathcal{B}}(\boldsymbol{\mu}) \\ &\text{s.t.} && \boldsymbol{\mu}_{\min} \leq \boldsymbol{\mu} \leq \boldsymbol{\mu}_{\max} \\ & && \text{and the constraints (12), (13) and (14)} \end{aligned} \tag{17}$$

while the maximization of the largest partial amplitude is governed by the optimization problem

$$\begin{aligned} &\underset{\boldsymbol{\mu}}{\text{maximize}} && \mathcal{F}_{\Gamma}^{(n)}(\boldsymbol{\mu}) \doteq \max_{h=2,\dots,11} \max_{p=1,3,4,6} \Delta\omega^{(n)}_{h(h+1),\Gamma_p}(\boldsymbol{\mu}) \\ &\text{s.t.} && \boldsymbol{\mu}_{\min} \leq \boldsymbol{\mu} \leq \boldsymbol{\mu}_{\max} \\ & && \text{and the constraints (12), (13) and (14)} \end{aligned} \tag{18}$$

Again, the absence of full and partial band gaps is revealed by non-positive values of the respective objective functions at optimality. As minor remark, by changing the order of the maximization, the problems (15)-(18) can be also tackled by, first, independently maximizing each $h(h+1)$-th band gap relative amplitude and, second, selecting the largest of all the obtained maxima.

Since the optimization problems are strongly nonlinear, their exact solution $\boldsymbol{\mu}^\circ$ (*optimal solution*) can seldom be determined analytically, but can alternatively be detected by means of numerical methods. At worst, the largest value numerically detectable for the objective functions can be considered an acceptable, although approximate, solution $\boldsymbol{\mu}^*$ (*suboptimal solution*). In the specific problem under investigation, a discretization of the Brillouin zone $\mathcal{B}$ is also needed for the evaluation of the objective function. In particular, a uniform $\mathcal{B}$-discretization is adopted by defining a $14 \times 14$ point grid. Similarly, a $\Gamma$-discretization has been assumed by determining 30 points equally spaced along each segment $\Gamma_p$. The mesh fineness has been motivated by a compromise between the two opposite requirements of numerical accuracy, on the one hand, and computational feasibility, on the other hand. However, the $\boldsymbol{\mu}^*$-vector obtained numerically has been always verified to provide the same large band gap relative amplitude using either the original grid or mesh refinement (consisting in a $60 \times 60$ point $\mathcal{B}$-discretization and a 100 point $\Gamma_p$-discretization). Analogous considerations can be pointed out for the $\boldsymbol{\mu_s}$-solutions. Moreover, it is worth noting that each discretization can be proved to always provide an upper bound on the objective function (see the Appendix C). Hence, if no band gaps are obtained for a particular choice of the grid fineness, it is very unlikely that a band gap is detected when a finer grid is adopted. This argument can be employed to justify a rapid check of the band gap absence, using even a coarse grid.

### *3.2. Numerical optimization methods*

The optimization problems have been solved by using the Globally Convergent version (GCMMA) [49] of the iterative Method of Moving Asymptotes (MMA) [50], combined with several initializations governed by a quasi-Monte Carlo multi-start technique. The term *moving asymptotes* means that the objective and constraint functions are approximated using functions whose asymptotes change from each iteration of the MMA to the successive one. The GCMMA version is *globally convergent* because it always converges to a stationary point, regardless of the choice of its initialization. Loosely speaking, the combination of the GCMMA and the quasi-Monte Carlo multi-start technique consists in tackling a sequence of concave-maximization subproblems, locally approximating the original nonlinear optimization problem (a different approximation at each sequence iteration). With respect to a single initialization, the Monte Carlo multi-start technique increases the probability of finding a global maximum point through a set of random initializations of the sequence. Furthermore, the choice of the quasi-Monte Carlo multi-start technique ensures that quasi-random initializations provide results similar to random initializations, but with the advantage of a better reproducibility. For the specific problem under investigation, the i-th initialization ($i = 1, ..., m$) furnishes the solution $\boldsymbol{\mu}_i^*$ corresponding to the local maximum $\mathcal{F}_i^* \doteq \mathcal{F}(\boldsymbol{\mu}_i^*)$. Therefrom, the suboptimal solution $\boldsymbol{\mu}^*$ is selected as that providing the largest $\mathcal{F}_i^*$, that is, the objective function value $\mathcal{F}^* \doteq \mathcal{F}(\boldsymbol{\mu}^*) = \max_{i=1,...,m} \mathcal{F}_i^*$. More details about the method are reported in [35]. It is worth remarking that both the MMA and GCMMA methods are often used to solve structural optimization problems [42, 50, 51].

An alternative approach, based on sequential linear programming with an adaptively-sized trust region

[52], could be considered. However, its application to the optimization problems under investigation has been verified to perform worse than the GCMMA, due to a slower convergence rate. Another alternative could be based on a sequence of semidefinite programming problems approximating the original problem [53]. Nonetheless, its application would be inconsistent (at least, not directly in the original form), due to the nonlinear dependence of some constraints on the optimization variables in the problems under investigation.

### *3.3. First optimization scenario*

In the first optimization scenario, all the problems stated in Section 3.1 have been tackled. The quasi-Montecarlo initialization of each problem required, first, to generate a sufficiently long quasi-random Sobol' sequence [54] in the normalized parameter unit cube and, second, to retain a shorter sequence (namely a *subsequence* of $m = 100$ points) satisfying the geometric constraints (12)-(14). Each point of this subsequence is employed to initialize a different repetition of the GCMMA. Therefore, 100 repetitions are carried out, each including a fixed number of $N = 30$ iterations. The method convergence after all the iterations is successfully verified a posteriori for the largest majority of the initializations.

The optimization results are summarized in Table 2, which provides the component values of the suboptimal $\boldsymbol{\mu}^*$-solution, and the corresponding objective value $\mathcal{F}^*$ for all the Cases $\mathcal{C}_{0-4}$. For the Cases $\mathcal{C}_0$ and $\mathcal{C}_4$, no full band gaps are obtained. Hence, the suboptimal solutions refer to partial band gaps. For all the other problems, instead, the suboptimal solutions refer to full band gaps. In addition, Figures 4-9 show the evolution of the objective value during the iterations of the GCMMA for each quasi-Monte Carlo initialization (Figures 4a-9a), the lowest 12 dispersion curves for the suboptimal solution (Figures 4b-9b). Furthermore the lowest 12 dispersion surfaces in the case of full band gap detection are shown (Figures 5b-8c), together with the $(h, k)$-th pair of dispersion surfaces associated with the detected band gap (Figures 5d,e,f-8d,e,f).

For the Case $\mathcal{C}_0$, the material spectrum systematically shows high density throughout the admissible region $\mathcal{R}_s$ spanned by the parameter subvector $\boldsymbol{\mu}_s$. Consequently, some curves are superimposed to each others, and many crossing phenomena occur among the spectrum branches. The subptimal solution identifies only a partial band gap located between the 4-th and 5-th dispersion curves along the segment $\mathit{\Gamma}_3$ (or $\mathit{\Gamma}_6$),

Table 2: First optimization scenario: component values of the suboptimal $\boldsymbol{\mu}^*$-solution and objective values $\mathcal{F}^*$.

| Case | | Optimization parameters | | | | | | | Objective and largest amplitude | | |
|---|---|---|---|---|---|---|---|---|---|---|---|
| $(n)$ | Fig. | $\frac{w_s}{\varepsilon}$ | $\frac{W}{w_s}$ | $\frac{R}{\varepsilon}$ | $\frac{R_r}{\varepsilon}$ | $\frac{E_r}{E_s}$ | $\nu_r$ | $\frac{\varrho_r}{\varrho}$ | function | value | amplitude |
| (0) | 4 | max | 1.0000 | max | - | - | - | - | $\mathcal{F}_{\mathit{\Gamma}}^{(0)}$ | 0.1740 | $\mathit{\Delta}\omega_{45}^{(0)}$ |
| (1) | 5 | 0.0632 | 0.1592 | max | 0.0684 | min | max | max | $\mathcal{F}_{\mathcal{B}}^{(1)}$ | 0.0934 | $\mathit{\Delta}\omega_{1112}^{(1)}$ |
| $(2h)$ | 6 | 0.0364 | 0.2749 | max | 0.0821 | min | max | max | $\mathcal{F}_{\mathcal{B}}^{(2h)}$ | 0.2238 | $\mathit{\Delta}\omega_{1011}^{(2h)}$ |
| $(2d)$ | 7 | 0.0592 | 0.1690 | max | 0.0824 | min | max | max | $\mathcal{F}_{\mathcal{B}}^{(2d)}$ | 0.3321 | $\mathit{\Delta}\omega_{67}^{(2d)}$ |
| (3) | 8 | 0.0700 | 0.1429 | max | 0.0822 | min | max | max | $\mathcal{F}_{\mathcal{B}}^{(3)}$ | 0.3086 | $\mathit{\Delta}\omega_{1112}^{(3)}$ |
| (4) | 9 | max | 1.0000 | max | 0.0420 | 0.9999 | 0.3991 | 1.2742 | $\mathcal{F}_{\mathit{\Gamma}}^{(4)}$ | 0.1690 | $\mathit{\Delta}\omega_{45}^{(4)}$ |

Legend: "min" and "max" stand for the minimum or maximum values reported in Table 1

which corresponds to acoustic waves propagating in the diagonal direction of the $\mathcal{B}$-space. In particular, the identified band gap is originated by a veering phenomenon and its relative amplitude is $\Delta\omega^{(0)}_{45,\Gamma_{3,6}} = 0.1740$.

For the Case $\mathcal{C}_1$, the meta-material spectrum again shows high density and frequency crossings throughout the admissible region $\mathcal{R}$ spanned by the parameter vector $\boldsymbol{\mu}$. Nonetheless, the subptimal solution identifies a full band gap between the 11-th and 12-th dispersion surfaces (Figure 5c,d), with relative amplitude $\Delta\omega^{(1)}_{1112,\mathcal{B}} = 0.0934$ (Figure 5a). It is worth remarking that the maximum of the 11-th surface and the minimum of the 12-th surface lie on two vertices of the $\Gamma$-curve (Figure 5e,f). As a consequence, the same full band gap could be detected by limiting the analysis to the dispersion branches along the $\Gamma$-curve (Figure 5b). As minor remark, a smaller full band gap shows up between the 4th and 5th dispersion surfaces.

For the Case $\mathcal{C}_{2h}$, the subptimal solution identifies a full band gap located between the 10-th and 11-th dispersion surfaces (Figure 6c,d), with relative amplitude $\Delta\omega^{(1)}_{1112,\mathcal{B}} = 0.2238$ (Figure 6a). Again, the same full band gap could be detected by limiting the analysis to the dispersion branches along the $\Gamma$-curve (Figure 6b). Smaller partial band gaps show up between the 6th and 7th dispersion curves, along the segments $\Gamma_1, \Gamma_3, \Gamma_4$, and $\Gamma_6$, corresponding to waves propagating in the horizontal, vertical and diagonal direction of the $\mathcal{B}$-space. For the Case $\mathcal{C}_{2d}$, the subptimal solution identifies a full band gap located between the 6-th and 7-th dispersion surfaces (Figure 7c,d), with relative amplitude $\Delta\omega^{(1)}_{67,\mathcal{B}} = 0.3321$ (Figure 7a). The other full band gaps (at lower frequencies between the 4-th and 5-th surfaces and at higher frequencies between the 10-th and 11-th surfaces) have smaller relative amplitude. All these band gaps could be detected by the analysis of the dispersion branches along the $\Gamma$-curve (Figure 7b).

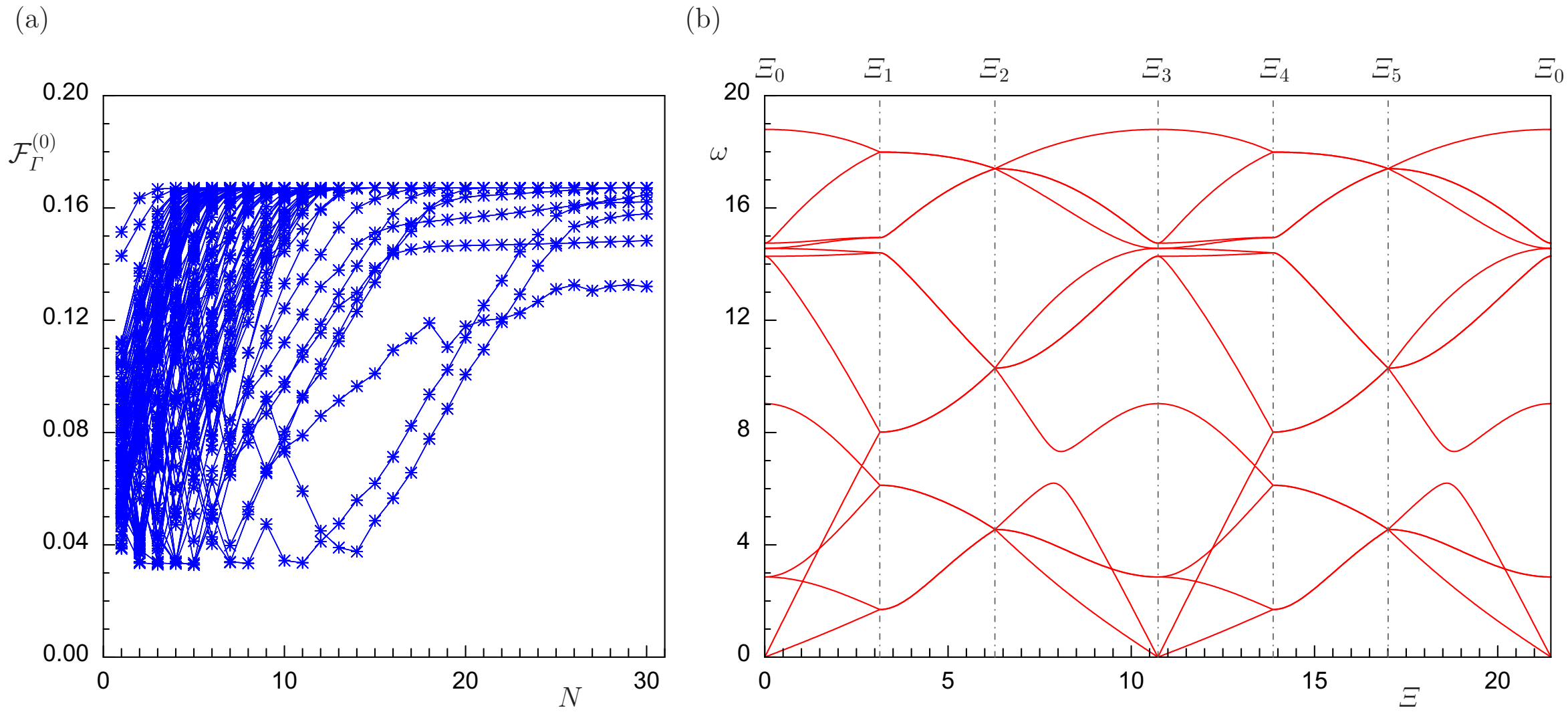

Figure 4: First optimization scenario for the Case $\mathcal{C}_0$: (a) converging objective values vs the iteration number for different GCMMA repetitions, (b) dispersion curves of the optimized spectrum.

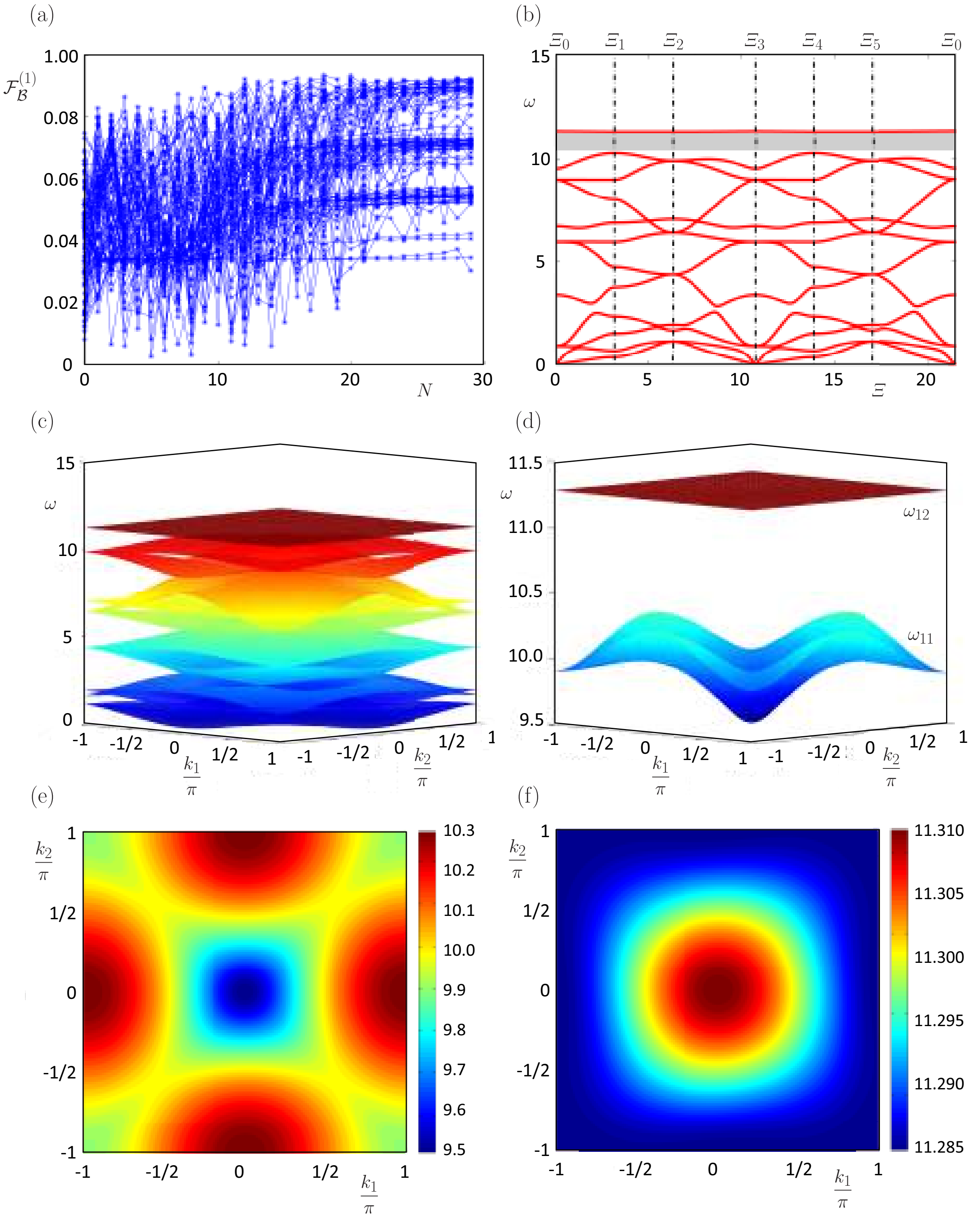

Figure 5: First optimization scenario for the Case $\mathcal{C}_1$: (a) converging objective values vs the iteration number for different GCMMA repetitions, (b),(c) dispersion curves and surfaces of the optimized spectrum, (d) optimal band gap, (e),(f) 11th and 12th dispersion surfaces in the $\mathcal{B}$-space.

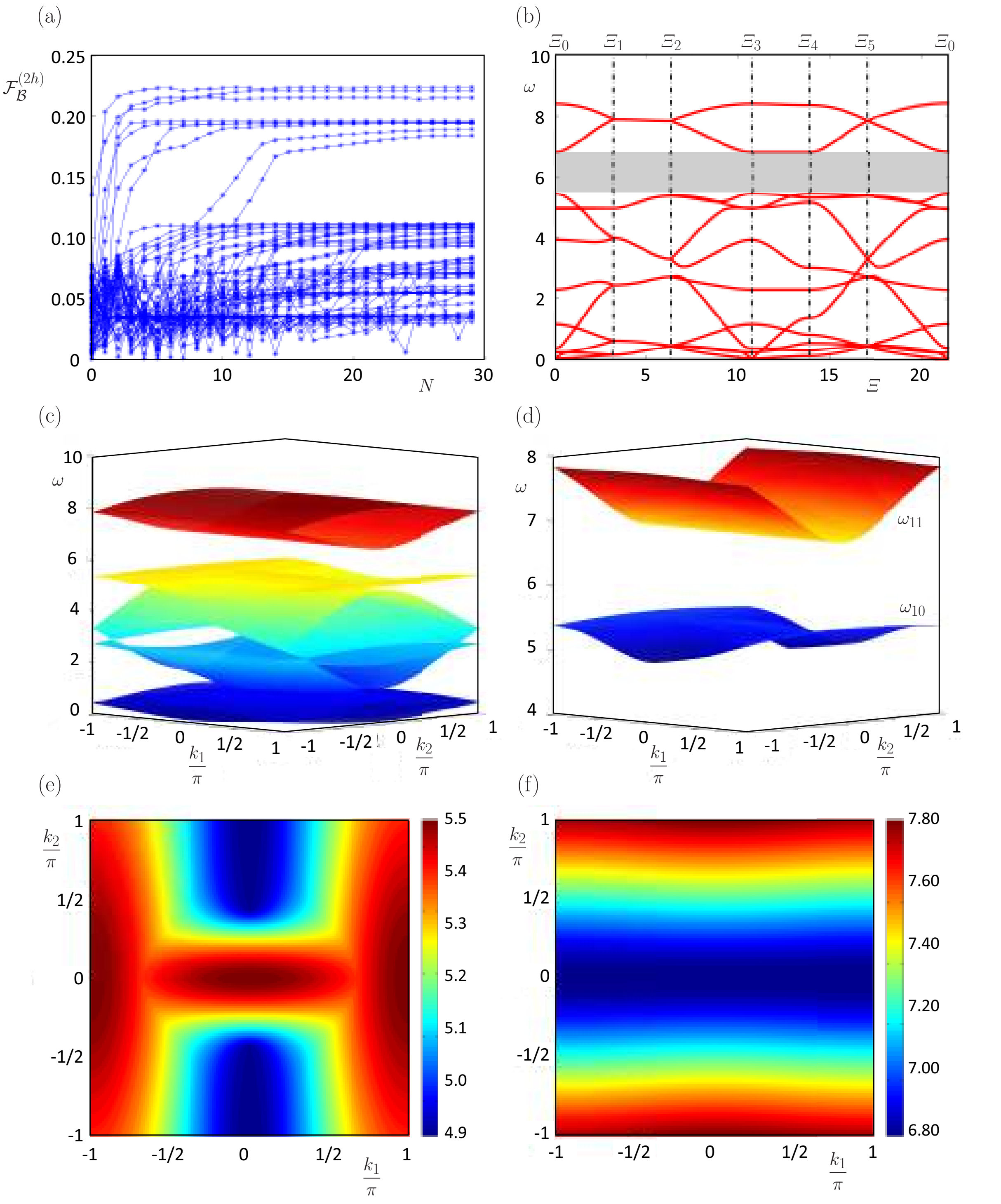

Figure 6: First optimization scenario for the Case $\mathcal{C}_{2h}$: (a) converging objective values vs the iteration number for different GCMMA repetitions, (b),(c) dispersion curves and surfaces of the optimized spectrum, (d) optimal band gap, (e),(f) 10th and 11th dispersion surfaces in the $\mathcal{B}$-space.

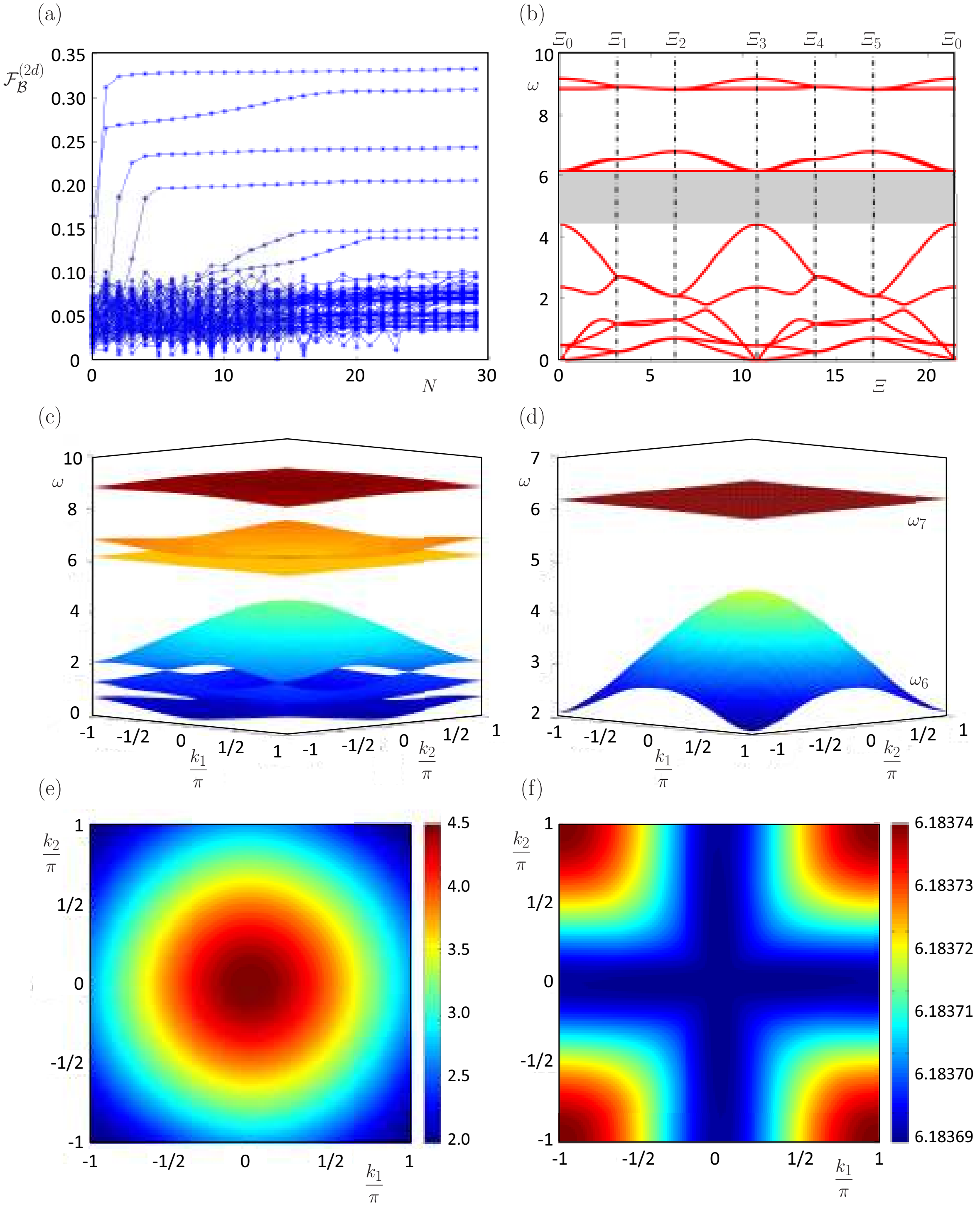

Figure 7: First optimization scenario for the Case $\mathcal{C}_{2d}$: (a) converging objective values vs the iteration number for different GCMMA repetitions, (b),(c) dispersion curves and surfaces of the optimized spectrum, (d) optimal band gap, (e),(f) 6th and 7th dispersion surfaces in the $\mathcal{B}$-space.

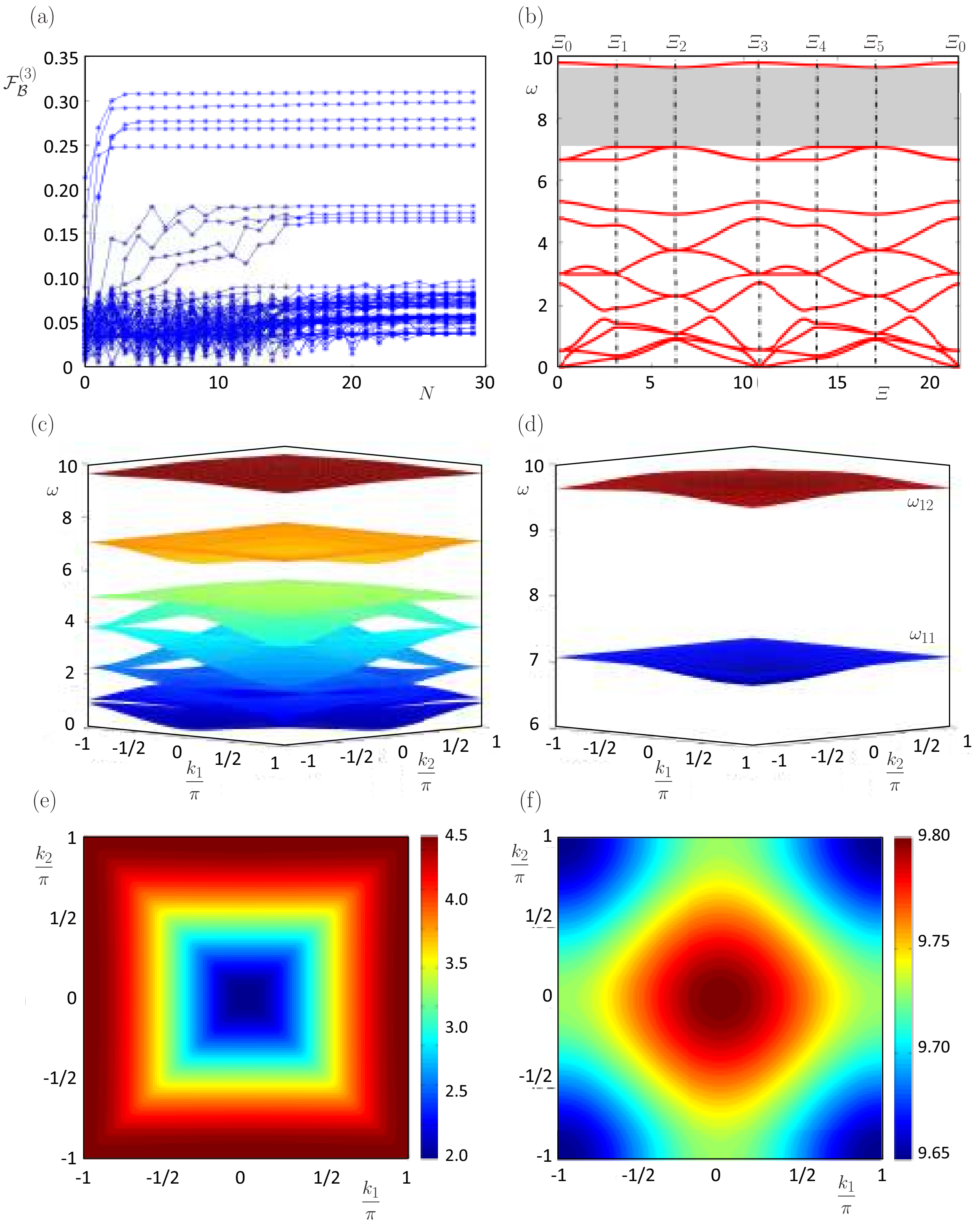

Figure 8: First optimization scenario for the Case $\mathcal{C}_3$: (a) converging objective values vs the iteration number for different GCMMA repetitions, (b),(c) dispersion curves and surfaces of the optimized spectrum, (d) optimal band gap, (e),(f) 11th and 12th dispersion surfaces in the $\mathcal{B}$-space.

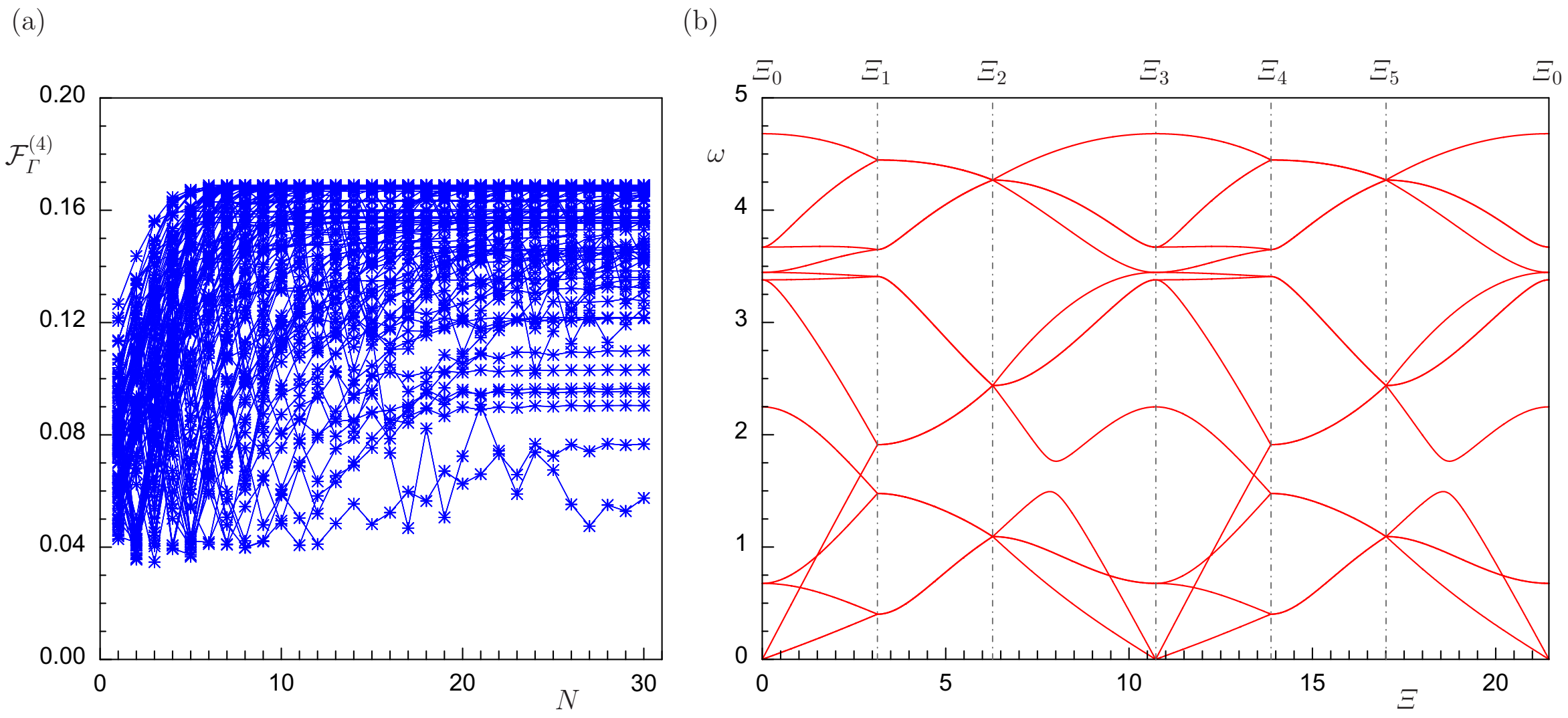

Figure 9: First optimization scenario for the Case $\mathcal{C}_4$: (a) converging objective values vs the iteration number for different GCMMA repetitions, (b) dispersion curves of the optimized spectrum.

For the Case $\mathcal{C}_3$, the subptimal solution identifies a full band gap located between the 11-th and 12-th dispersion surfaces (Figure 8c,d), with relative amplitude $\Delta\omega^{(1)}_{1112,\mathcal{B}} = 0.3086$ (Figure 8a). Three other band gaps can be observed at lower frequencies, with smaller absolute and relative amplitudes. Again, all these band gaps could be detected by the analysis of the dispersion branches along the $\Gamma$-curve (Figure 8b). Finally, for the Case $\mathcal{C}_4$, the subptimal solution identifies a full band gap located between the 4-th and 5-th dispersion surfaces (Figure 9c,d), with relative amplitude $\Delta\omega^{(1)}_{67,\mathcal{B}} = 0.1690$ (Figure 9a). Furthermore, a partial band gap located between the 5-th and 6-th dispersion curves can be detected along the segment $\Gamma_3$ (or $\Gamma_6$), which corresponds to acoustic waves propagating in the diagonal direction of the $\mathcal{B}$-space.

Comparing all the material and meta-material configurations, the highest full amplitude is achievable in the Case $\mathcal{C}_{2d}$ (*best* case), corresponding to the cellular configuration of the meta-material equipped with two inertial resonators housed by the rings along one of the square cell diagonals. Sorting the Cases in descending order of the objective value, the Case $\mathcal{C}_{2d}$ is followed by the Cases $\mathcal{C}_3$, $\mathcal{C}_{2h}$, $\mathcal{C}_0$, $\mathcal{C}_4$, $\mathcal{C}_1$. Among all the suboptimal solutions, the highest full amplitude obtained in the best Case $\mathcal{C}_{2d}$ is also referred to the band gap between the lowest pair of dispersion surfaces (the 6th and 7th). It is worth mentioning that the assessment of the full amplitude is robust in the neighborhood of the suboptimal solutions $\boldsymbol{\mu}^*$ [55, 56]. Indeed, adding the perturbation $\boldsymbol{\eta}$ to the vector $\boldsymbol{\mu}^*$, the perturbed full amplitude can be demonstrated to be approximable by $\mathcal{F}^*_{\mathcal{B}} + \|\boldsymbol{\eta}\|_2 \left\| \left(\nabla \mathcal{F}_{\mathcal{B}}(\boldsymbol{\eta})\right)_{\boldsymbol{\eta}=\mathbf{0}} \right\|_2$, where $\|\cdot\|_2$ is the Euclidean norm on $\mathbb{R}^7$ and the second term comes from bounding from above the Euclidean norm of the perturbation, and applying the Cauchy-Schwarz inequality to the inner product between two vectors. For instance, fixing the perturbation $\boldsymbol{\eta} = \eta\mathbf{1}$ with $\eta = 0.1$ and unitary vector $\mathbf{1}$, the full amplitude $\Delta\omega^{(2d)}_{67,\mathcal{B}}$ decreases from 0.3321 (suboptimal value) to 0.3152.

Quite surprisingly, no full band gaps are obtained for the Case $\mathcal{C}_4$, that is, for the maximum number of resonators. However, this cannot be considered a contradiction, since the Case $\mathcal{C}_4$ differs from the other Cases for the addiction of one or more resonators with identical geometrical and mechanical properties. Indeed, complementary results (here not reported for the sake of synthesis) show that large amplitudes of full band gap can be achieved by independently optimizing the properties of each of four non-identical resonators. Differently, when at least one full band gap is discovered (Cases $\mathcal{C}_{1-3}$), the meta-material spectrum often exhibits one or more smaller band gaps, for different pairs of surfaces.

It is also worth mentioning that many of the parameters attain the same suboptimal values in one or more Cases, if these suboptimal values coincide with the maximum or minimum bounds of the $\boldsymbol{\mu}$-vector. Interestingly, all the full amplitudes obtained as suboptimal solutions of the problem (17), that is, coming out from the maximization extended to the full $\mathcal{B}$-space, have been systematically confirmed by the suboptimal solutions obtainable by formulating and solving the optimization problem

$$\begin{aligned} \underset{\boldsymbol{\mu}}{\text{maximize}} \quad & \max_{h=2,\dots,11} \Delta\omega^{(n)}_{h(h+1),\Gamma}(\boldsymbol{\mu}) \\ \text{s.t.} \quad & \boldsymbol{\mu}_{\min} \leq \boldsymbol{\mu} \leq \boldsymbol{\mu}_{\max} \\ & \text{and the constraints (12), (13) and (14)} \end{aligned} \tag{19}$$

which differs from the problem (17) for the maximization limited to the closed $\Gamma$-curve.

From the physical viewpoint, some recurrent findings can definitely be recognized as systematic trends for the meta-material optimization. According to the particular selection of the performance criteria and with the appropriate caution in the numerical result extrapolation, the anti-tetrachiral meta-materials have been found to highly perform when they are, first, characterized by a strong ring-to-ligament compositeness of the cell microstructure (maximum geometric ratio $R/\varepsilon$) and, second, equipped with a few large and heavy resonators (namely $R_r \simeq 0.82R$ and maximum inertial ratio $\varrho_r/\varrho$) weakly coupled with their hosting rings (minimum elastic ratio $E_r/E$). More than two identical resonators tend to reduce the meta-material performance, whereas the diagonal placement of two resonators allows a twofold achievement, that is, the largest amplitude of a full band gap for the lowest pair of consecutive dispersion surfaces.

### *3.4. Second optimization scenario*

To strengthen the findings of the first optimization scenario, a second-level numerical analysis is carried out with the aim to further enhance the performance of the meta-materials already exhibiting the highest full amplitudes. To this purposes, a second optimization effort is focused on the Cases $\mathcal{C}_{1-3}$, by preliminary fixing the microstructural parameters according to the suboptimal solution of the best Case $\mathcal{C}_{2d}$ (namely $w_s/\varepsilon = 0.0592, W/w_s = 0.1690, R/\varepsilon = 0.1000$, see Table 2). Therefore, only the resonator properties $(R_r/\varepsilon, E_r/E_s, \nu_r, \varrho_r/\varrho)$ are considered as second-level optimization variables, and the optimization problems (17),(18) have been modified accordingly. Moreover, a smaller number of admissible repetitions of the GCMMA ($m = 25$) has been found sufficient to deal with a lower number of optimizable parameters.

The achievements of the new numerical analyses are summarized in Table 3, which provides the component values of the new (partially fixed) suboptimal $\boldsymbol{\mu}^*$-solution, and the corresponding objective value $\mathcal{F}^*$ for the Cases $\mathcal{C}_{1-3}$. In addition, Figures 10-12 show the evolution of the objective value during the iterations of the

Table 3: Second optimization scenario: component values of the suboptimal $\boldsymbol{\mu}^*$-solution and objective values $\mathcal{F}^*$.

| Case | | Optimization parameters | | | | | | | Objective and largest amplitude | | |
|---|---|---|---|---|---|---|---|---|---|---|---|
| $(n)$ | Fig. | $\frac{w_s}{\varepsilon}$ | $\frac{W}{w_s}$ | $\frac{R}{\varepsilon}$ | $\frac{R_r}{\varepsilon}$ | $\frac{E_r}{E_s}$ | $\nu_r$ | $\frac{\varrho_r}{\varrho}$ | function | value | amplitude |
| (1) | 10 | = | = | = | 0.0668 | min | max | max | $\mathcal{F}_{\mathcal{B}}^{(1)}$ | 0.0927 | $\Delta\omega_{1112}^{(1)}$ |
| $(2h)$ | 11 | = | = | = | 0.0717 | min | max | max | $\mathcal{F}_{\mathcal{B}}^{(2h)}$ | 0.2012 | $\Delta\omega_{1011}^{(2h)}$ |
| $(2d)$ | 12 | 0.0592 | 0.1690 | 0.1000 | 0.0827 | min | max | max | $\mathcal{F}_{\mathcal{B}}^{(2d)}$ | 0.3350 | $\Delta\omega_{67}^{(2d)}$ |
| (3) | 13 | = | = | = | 0.0826 | min | max | max | $\mathcal{F}_{\mathcal{B}}^{(3)}$ | 0.2886 | $\Delta\omega_{1112}^{(3)}$ |

Legend: "=" stands for equal to Case $(n)=(2d)$, "min" and "max" stand for the minimum or maximum value in Table 1

GCMMA for each repetition (Figures 10a-12a), the lowest 12 dispersion curves for the suboptimal solution (Figures 10b-12b). Furthermore the lowest 12 dispersion surfaces are shown (Figures 10b-12c), together with the $(h,k)$-th pair of dispersion surfaces associated with the detected band gap (Figures 10d,e,f-12d,e,f).

For the Case $\mathcal{C}_1$, fixing the microstructural parameters implies minor differences in the $w_s/\varepsilon$-value (+6.3%) and the $W/w_s$-value (-6.2%) with respect to the *first suboptimal solution* (rigorously, the suboptimal solution of the first optimization scenario). In the *second suboptimal solution* (rigorously, the suboptimal solution of the second-level optimization), the full band gap between the 11-th and 12-th dispersion surfaces (Figure 10c,d) is still present, although with a slight decrement (-0.75%) of the relative amplitude $\Delta\omega_{1112,\mathcal{B}}^{(1)}=0.0927$ (Figure 10a). Again, the same full band gap could be also detected by limiting the analysis to the dispersion branches along the $\Gamma$-curve (Figure 10b). The smaller full band gap between the 4-th and 5-th dispersion surfaces is reduced in the amplitude but not cancelled by the second optimization process.

For the Case $\mathcal{C}_{2h}$, fixing the microstructural parameters implies strong differences in the $w_s/\varepsilon$-value (-62.6%) and the $W/w_s$-value (+38.5%) with respect to the first suboptimal solution. Nonetheless, the full band gap between the 10-th and 11-th dispersion surfaces (Figure 11c,d) still exists in the second suboptimal solution, although centered at higher frequencies and with significant decrement (-10.1%) of the relative amplitude $\Delta\omega_{1011,\mathcal{B}}^{(2h)}=0.2012$ (Figure 11a). Apart this high-frequency band gap, no other full band gaps can be detected by limiting the analysis to the dispersion branches along the $\Gamma$-curve (Figure 11b).

Considering the fixed parameters, the second suboptimal solution of the best Case $\mathcal{C}_{2d}$ is essentially characterized by a small increment of the $R_r/\varepsilon$-value (+0.36%) with respect to the first optimization scenario. The full band gap between the 6-th and 7-th dispersion surfaces (Figure 11c,d) is slightly enlarged, with an increment (+0.87%) of the relative amplitude $\Delta\omega_{67,\mathcal{B}}^{(1)}=0.3350$ (Figure 12a). Smaller full band gaps are still characterizing the lower frequency range (between the 4-th and 5-th surfaces) and the higher frequency range (between the 10-th and 11-th surfaces), with nearly-unchanged amplitudes.

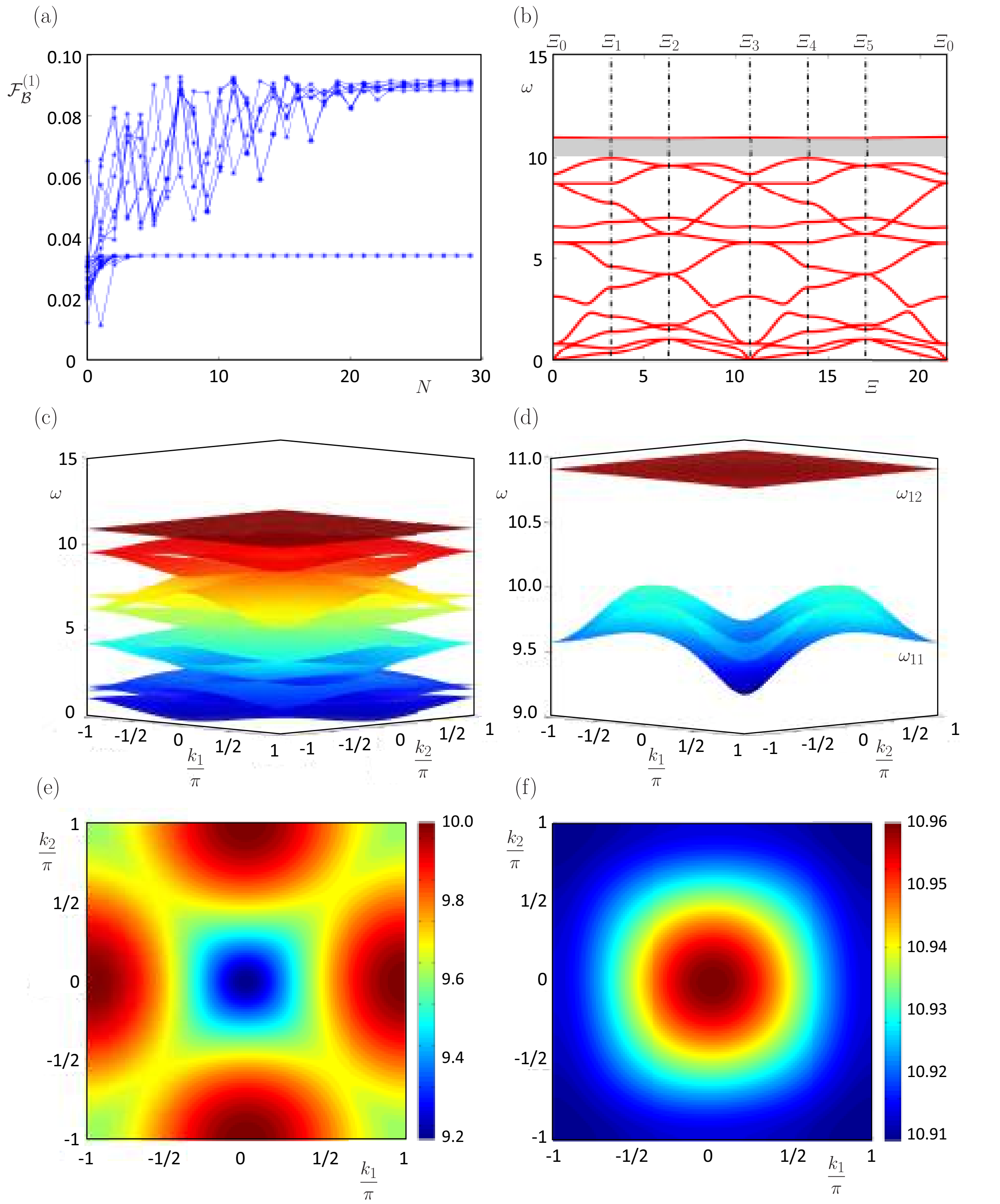

Figure 10: Second optimization scenario for the Case $\mathcal{C}_1$: (a) converging objective values vs the iteration number for different GCMMA repetitions, (b),(c) dispersion curves and surfaces of the optimized spectrum, (d) optimal band gap, (e),(f) 11th and 12th dispersion surfaces in the $\mathcal{B}$-space.

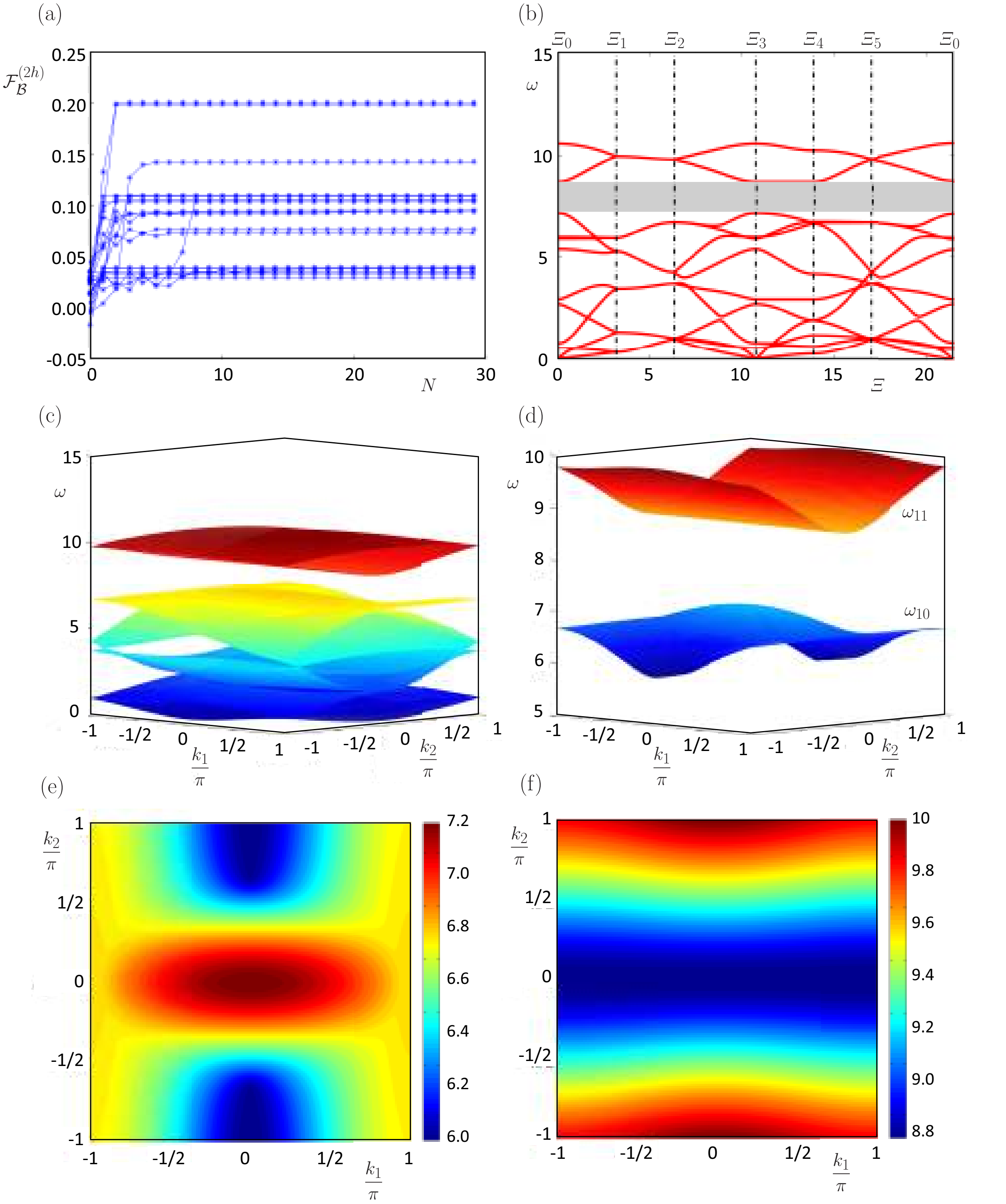

Figure 11: Second optimization scenario for the Case $\mathcal{C}_{2h}$: (a) converging objective values vs the iteration number for different GCMMA repetitions, (b),(c) dispersion curves and surfaces of the optimized spectrum, (d) optimal band gap, (e),(f) 10th and 11th dispersion surfaces in the $\mathcal{B}$-space.

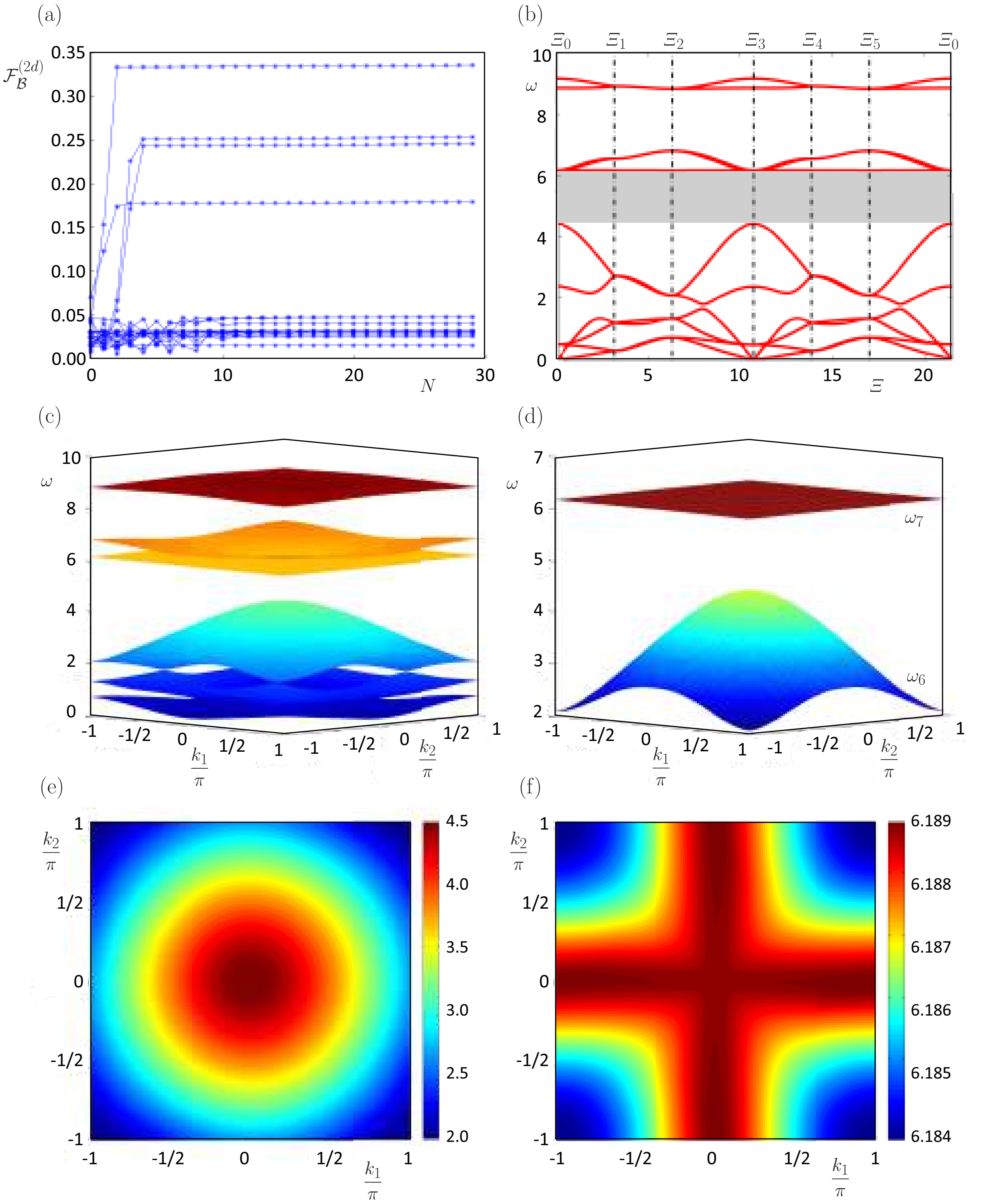

Figure 12: Second optimization scenario for the Case $\mathcal{C}_{2d}$: (a) converging objective values vs the iteration number for different GCMMA repetitions, (b),(c) dispersion curves and surfaces of the optimized spectrum, (d) optimal band gap, (e),(f) 6th and 7th dispersion surfaces in the $\mathcal{B}$-space.

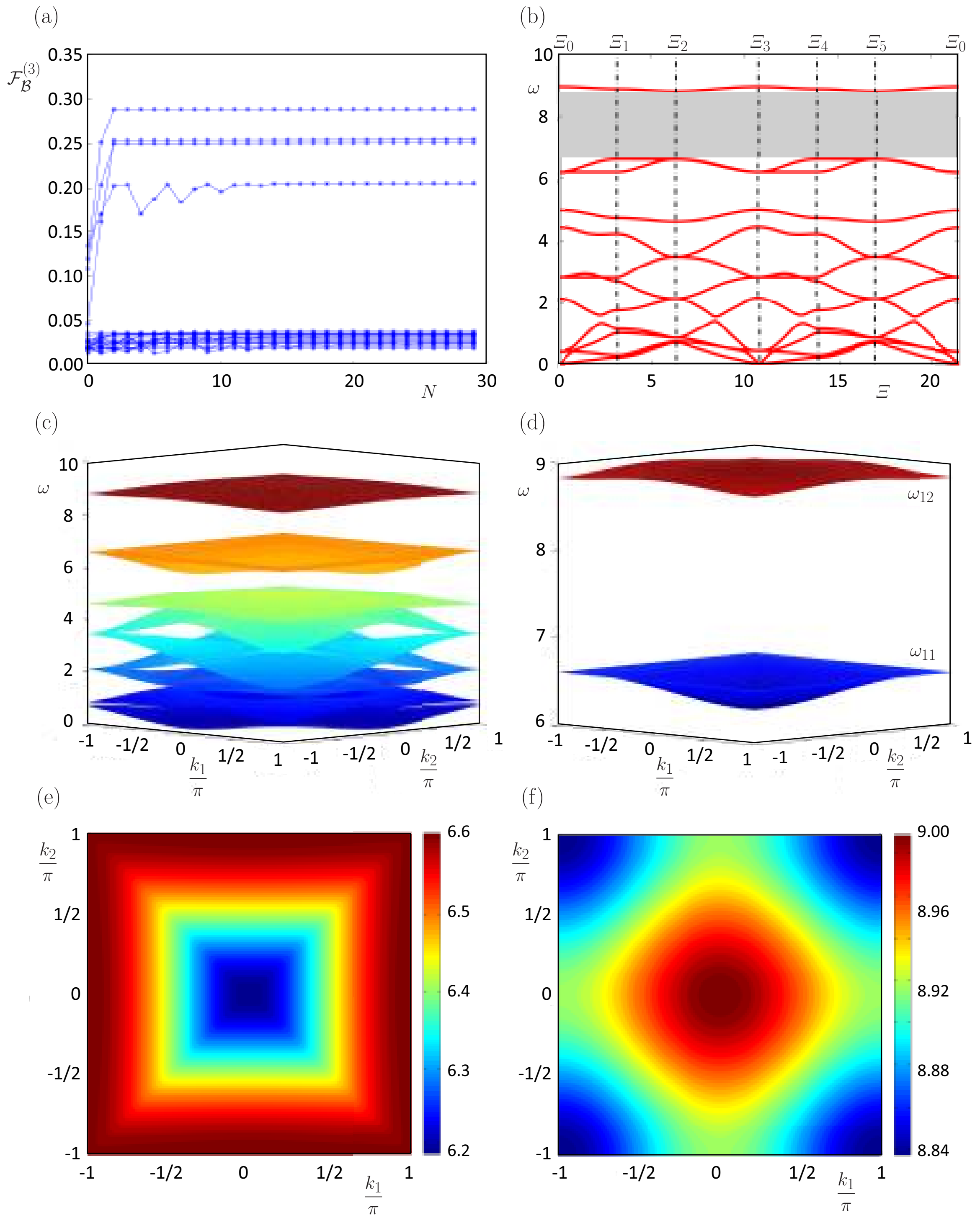

Figure 13: Second optimization scenario for the Case $\mathcal{C}_3$: (a) converging objective values vs the iteration number for different GCMMA repetitions, (b),(c) dispersion curves and surfaces of the optimized spectrum, (d) optimal band gap, (e),(f) 11th and 12th dispersion surfaces in the $\mathcal{B}$-space.

For the Case $\mathcal{C}_1$, fixing the microstructural parameters entails significant modification the $w_s/\varepsilon$-value (15.6%) and the $W/w_s$-value (-18.5%) with respect to the first suboptimal solution. In the second suboptimal solution, the full band gap between the 11-th and 12-th dispersion surfaces (Figure 13c,d) still exits, but centered at lower frequencies and with significant decrement (-6.48%) of the relative amplitude $\Delta\omega^{(1)}_{1112,\mathcal{B}} = 0.2886$ (Figure 13a). All the other full band gaps with smaller amplitudes shift to lower frequency ranges.

In summary, the second optimization provides a minor performance enhancement of the meta-material equipped with two diagonal resonators, by virtue of a slight increment of the resonator radius. On the contrary, the second optimization systematically degrades the performance of the other meta-material configurations. Indeed, their second suboptimal solutions remarkably suffer from the reduction in the dimension of free parameter space, together with the imposition of non-optimal fixed parameters.

## 4. Conclusions

A linear dynamic model has been formulated to analytically describe to wave propagation properties of a composite lattice material characterized by a square periodic cell with anti-tetrachiral topology. An inertial meta-material has been created through the introduction of local resonators in the cellular microstructure, composed by a regular pattern of rigid rings and flexible ligaments. The consequent increment of the model dimension has strongly modified the material band structure, already featured by a marked spectral density. By virtue of the elastic ring-resonator coupling, strong linear interactions among the dispersion curves may occur. This powerful fully-mechanical effect has been exploited to improve the meta-material performance as tailor-made passive filter for low-frequency elastic waves, by properly tuning the mechanical parameters.

Starting from the direct linear eigenproblem governing the wave propagation problem for periodic system, a nonlinear optimization problem has been formulated by adopting the largest band gap amplitude in the low-frequency meta-material spectrum as objective function. The optimal solution which maximizes the objective function has been searched in a bounded region of the parameter space, satisfying the necessary constrain conditions of geometrical consistency and physical feasibility. Full and partial band gaps have been considered as primary and secondary optimization targets, respectively. Two independent subsets of mechanical parameters, one related to the material microstructure and the other concerning the elastic and inertial resonator properties, have been considered as optimization variables. Different cell configurations, with variable number and placement of the resonators, have been explored as design alternatives. The optimization problem has been tackled by a nonlinear numerical technique, featured by global convergence, accompanied by a quasi-random multi-start initialization.

From a qualitative viewpoint, the optimal solution has shown that the anti-tetrachiral material does not offer full low-frequency band gaps in the absence of resonators. Partial band gaps can be achieved for dispersive waves propagating along the diagonal directions of the square cell, which differ from the material orthotropy axes. Similarly, partial but not full band gaps are achievable in the anti-tetrachiral meta-material equipped with the maximum number of identical resonators. On the contrary, a lower number of resonators always allows the opening of one or more full band gaps in the low-frequency range.

From the physical viewpoint, the anti-tetrachiral meta-materials have been systematically found to offer the largest band gap amplitudes when they are, first, characterized by a strong ring-to-ligament compositeness of the cell microstructure (maximum admissible ring radius) and, second, equipped with a few large and heavy resonators (maximum admissible resonator inertia) weakly coupled with their hosting rings (minimum elastic resonator stiffness). More than two identical resonators tend to reduce the meta-material performance, whereas the diagonal placement of two resonators allows the simultaneous achievement of the largest full amplitude in the lowest frequency range. These findings have also shown a certain qualitative and quantitative robustness in respect to small perturbations of the optimal parameter combinations, even when a reduction of the free parameter space is forced in the optimization problem.

## Appendix A. Equations of motion

The active internal nodes develop both nondimensional elastic $\boldsymbol{\sigma}_a = (\boldsymbol{\sigma}_s, \boldsymbol{\sigma}_r)$ and inertial forces $\mathbf{f}_a = (\mathbf{f}_s, \mathbf{f}_r)$, which actively participate in the dynamic cell equilibrium. On the contrary, the passive external nodes can develop only elastic forces $\boldsymbol{\sigma}_p$, which partially depend on the stiffness coupling with the internal nodes, and quasi-statically balance the reactive forces $\mathbf{f}_p$ transferred by the adjacent cells.

According to displacement/force decomposition, the nondimensional equilibrium equation governing the undamped free oscillations of the discrete model has the vector form

$$\begin{pmatrix} \mathbf{f}_s \\ \mathbf{f}_r \\ \mathbf{0} \end{pmatrix} + \begin{pmatrix} \boldsymbol{\sigma}_s \\ \boldsymbol{\sigma}_r \\ \boldsymbol{\sigma}_p \end{pmatrix} = \begin{pmatrix} \mathbf{0} \\ \mathbf{0} \\ \mathbf{f}_p \end{pmatrix} \tag{A.1}$$

or, making explicit the force dependence on the nodal acceleration or displacements

$$\begin{bmatrix} \mathbf{M}_s & \mathbf{O} & \mathbf{O} \\ \mathbf{O} & \mathbf{M}_r & \mathbf{O} \\ \mathbf{O} & \mathbf{O} & \mathbf{O} \end{bmatrix} \begin{pmatrix} \ddot{\mathbf{q}}_s \\ \ddot{\mathbf{q}}_r \\ \ddot{\mathbf{q}}_p \end{pmatrix} + \begin{bmatrix} \mathbf{K}_{ss} + \mathbf{K}_r & -\mathbf{K}_r & \mathbf{K}_{sp} \\ -\mathbf{K}_r & \mathbf{K}_r & \mathbf{O} \\ \mathbf{K}_{ps} & \mathbf{O} & \mathbf{K}_{pp} \end{bmatrix} \begin{pmatrix} \mathbf{q}_s \\ \mathbf{q}_r \\ \mathbf{q}_p \end{pmatrix} = \begin{pmatrix} \mathbf{0} \\ \mathbf{0} \\ \mathbf{f}_p \end{pmatrix} \tag{A.2}$$

where dot indicates differentiation with respect to the $\tau$-time and $\mathbf{O}$ stands for matrices with all-zero elements.

Focusing on the micro-structure, the mass submatrix $\mathbf{M}_s$ is diagonal, as far as a lumped mass description is assumed. The symmetric submatrices $\mathbf{K}_{ss}$ and $\mathbf{K}_{pp}$ account for the stiffness of the internal and external nodes, respectively. The rectangular submatrix $\mathbf{K}_{sp} = \mathbf{K}_{ps}^\top$ expresses the elastic coupling among the internal and external nodes. Focusing on the resonators, both the local mass and stiffness submatrices $\mathbf{M}_r$ and $\mathbf{K}_r$ are diagonal. The submatrix $\mathbf{K}_r$ accounts also for the ring-resonator coupling in the internal nodes.

The free wave propagation along the bi-dimensional cell domain can be studied according to the Floquet-Bloch theory. Moving in the $\mathbf{k}$-transformed space the active ($j = 1...4, 13...16$) and passive displacements and passive force vectors assume the representations

$$\mathbf{q}_j = \tilde{\mathbf{q}}_j \exp\left(\iota \mathbf{k} \cdot \mathbf{x}_j\right), \qquad \mathbf{q}_p = \mathbf{F}_p \tilde{\mathbf{q}}_p, \qquad \mathbf{f}_p = \mathbf{F}_p \tilde{\mathbf{f}}_p \tag{A.3}$$

where $\iota$ denotes the imaginary unit, $\mathbf{k} = (k_1, k_2)$ is the (dimensional) wavevector and the block diagonal matrix $\mathbf{F}_p = \mathrm{diag}\left[\mathbf{I}\exp\left(\iota\mathbf{k}\cdot\mathbf{x}_5\right), ..., \mathbf{I}\exp\left(\iota\mathbf{k}\cdot\mathbf{x}_{12}\right)\right]$ with $\mathbf{I}$ being the three-by-three unit matrix.

The passive displacement and force vector can be ordered and partitioned as $\mathbf{q}_p = (\mathbf{q}_p^-, \mathbf{q}_p^+)$, $\mathbf{f}_p = (\mathbf{f}_p^-, \mathbf{f}_p^+)$ to separate the variable pairs $(\mathbf{q}_p^-, \mathbf{f}_p^-)$ belonging to the left/bottom cell boundary (composed by the external nodes ⑤,⑦,⑨,⑩) from the variable pairs $(\mathbf{q}_p^+, \mathbf{f}_p^+)$ belonging to the right/top boundary (composed by the external nodes ⑥,⑧,⑪,⑫). Extending the same partition to the respective transformed variables, the equation (A.3) can be written

$$\mathbf{q}_p^- = \mathbf{F}_p^-\tilde{\mathbf{q}}_p^-, \qquad \mathbf{q}_p^+ = \mathbf{F}_p^+\tilde{\mathbf{q}}_p^+, \qquad \mathbf{f}_p^- = \mathbf{F}_p^-\tilde{\mathbf{f}}_p^-, \qquad \mathbf{f}_p^+ = \mathbf{F}_p^+\tilde{\mathbf{f}}_p^+, \tag{A.4}$$

where, based on the decomposition above, the matrices $\mathbf{F}_p^-$ and $\mathbf{F}_p^+$ are defined as

$$\mathbf{F}_p^- = \mathrm{diag}\left[\mathbf{I}\exp\left(\iota\mathbf{k}\cdot\mathbf{x}_5\right), \mathbf{I}\exp\left(\iota\mathbf{k}\cdot\mathbf{x}_7\right), \mathbf{I}\exp\left(\iota\mathbf{k}\cdot\mathbf{x}_9\right), \mathbf{I}\exp\left(\iota\mathbf{k}\cdot\mathbf{x}_{10}\right)\right] \tag{A.5}$$

$$\mathbf{F}_p^+ = \mathrm{diag}\left[\mathbf{I}\exp\left(\iota\mathbf{k}\cdot\mathbf{x}_6\right), \mathbf{I}\exp\left(\iota\mathbf{k}\cdot\mathbf{x}_8\right), \mathbf{I}\exp\left(\iota\mathbf{k}\cdot\mathbf{x}_{11}\right), \mathbf{I}\exp\left(\iota\mathbf{k}\cdot\mathbf{x}_{12}\right)\right] \tag{A.6}$$

where $\mathbf{x}_j$ is the position vector pointing the $j$-th node.

Imposing the periodicity conditions on the transformed variables ($\tilde{\mathbf{q}}_p^+ = \tilde{\mathbf{q}}_p^-$ and $\tilde{\mathbf{f}}_p^+ = -\tilde{\mathbf{f}}_p^-$), the free wave propagation throughout the cell domain between the two complementary boundaries is governed by the quasi-periodicity conditions on the anti-transformed variables

$$\mathbf{q}_p^+ = \mathbf{L}\mathbf{q}_p^-, \qquad \mathbf{f}_p^+ = -\mathbf{L}\mathbf{f}_p^- \tag{A.7}$$

where, following from the equations (A.4), the block diagonal transfer matrix reads

$$\mathbf{L} = \mathrm{diag}\left[\exp\left(\iota\mathbf{k}\cdot\mathbf{d}_{56}\right)\mathbf{I}, \exp\left(\iota\mathbf{k}\cdot\mathbf{d}_{78}\right)\mathbf{I}, \exp\left(\iota\mathbf{k}\cdot\mathbf{d}_{911}\right)\mathbf{I}, \exp\left(\iota\mathbf{k}\cdot\mathbf{d}_{1012}\right)\mathbf{I}\right] \tag{A.8}$$

and $\mathbf{d}_{ij} = \mathbf{x}_j - \mathbf{x}_i$ represents the vector connecting the $i$-th and the $j$-th external nodes (Figure 2).

Consistently with the passive displacement and force decomposition, and imposing the quasi-periodicity conditions (A.7), the lower (quasi-static) part of equation (A.2) reads

$$\begin{bmatrix} \mathbf{K}_{ps}^- \\ \mathbf{K}_{ps}^+ \end{bmatrix} \mathbf{q}_s + \begin{bmatrix} \mathbf{K}_{pp}^= & \mathbf{K}_{pp}^\mp \\ \mathbf{K}_{pp}^\pm & \mathbf{K}_{pp}^\# \end{bmatrix} \begin{bmatrix} \mathbf{I} \\ \mathbf{L} \end{bmatrix} \mathbf{q}_p^- = \begin{bmatrix} \mathbf{I} \\ -\mathbf{L} \end{bmatrix} \mathbf{f}_p^- \tag{A.9}$$

with $\mathbf{I}$ being now the twelve-by-twelve unit matrix. This equation can be solved to express the passive variables as slave functions of the master active displacements, yielding

$$\mathbf{q}_p^- = \mathbf{R}\left(\mathbf{K}_{ps}^+ + \mathbf{L}\mathbf{K}_{ps}^-\right)\mathbf{q}_s, \qquad \mathbf{f}_p^- = \left(\mathbf{K}_{ps}^- + \left(\mathbf{K}_{pp}^= + \mathbf{K}_{pp}^\mp\mathbf{L}\right)\mathbf{R}\left(\mathbf{K}_{ps}^+ + \mathbf{L}\mathbf{K}_{ps}^-\right)\right)\mathbf{q}_s \tag{A.10}$$

where the $\mathbf{k}$-dependent matrix $\mathbf{R} = -\left(\mathbf{L}\mathbf{K}_{pp}^\mp\mathbf{L} + \mathbf{L}\mathbf{K}_{pp}^= + \mathbf{K}_{pp}^\#\mathbf{L} + \mathbf{K}_{pp}^\pm\right)^{-1}$ is diagonal.

Similarly, the imposition of the quasi-periodicity conditions to the upper (dynamic) part of the equation (A.2) leads to a coupled equation which, after condensation of the passive variables by virtue of the enslaving relations (A.10), depends on the active variables only

$$\mathbf{M}\ddot{\mathbf{q}}_a + \mathbf{K}\mathbf{q}_a = \begin{bmatrix} \mathbf{M}_s & \mathbf{O} \\ \mathbf{O} & \mathbf{M}_r \end{bmatrix} \begin{pmatrix} \ddot{\mathbf{q}}_s \\ \ddot{\mathbf{q}}_r \end{pmatrix} + \begin{bmatrix} \mathbf{K}_s & -\mathbf{K}_r \\ -\mathbf{K}_r & \mathbf{K}_r \end{bmatrix} \begin{pmatrix} \mathbf{q}_s \\ \mathbf{q}_r \end{pmatrix} = \begin{pmatrix} \mathbf{0} \\ \mathbf{0} \end{pmatrix} \tag{A.11}$$

where the condensed stiffness matrix $\mathbf{K}_s = \mathbf{K}_s^* + \mathbf{K}_r$ and $\mathbf{K}_s^* = \mathbf{K}_{ss} + (\mathbf{K}_{sp}^- + \mathbf{K}_{sp}^+ \mathbf{L})\mathbf{R}(\mathbf{K}_{ps}^+ + \mathbf{L}\mathbf{K}_{ps}^-)$, with the symmetries $\mathbf{K}_{sp}^- = (\mathbf{K}_{ps}^-)^\top$ and $\mathbf{K}_{sp}^+ = (\mathbf{K}_{ps}^+)^\top$ is known to be Hermitian.

## Appendix B. Parametric form of the mass and complex stiffness matrices

The non null elements of the 12-by-12 diagonal submatrix $\mathbf{M}_s$ (with components $M_{hk}^s$ and $h, k = 1, ..., 12$) and the 3-by-3 $j$-th ($j = 13...16$) diagonal submatrices $\mathbf{M}_j^r$ (with components $M_{hk}^r$ and $h, k = 1, 2, 3$) read

$$M_{1\,1}^s = M_{2\,2}^s = M_{4\,4}^s = M_{5\,5}^s = M_{7\,7}^s = M_{8\,8}^s = M_{10\,10}^s = M_{11\,11}^s = 2\,\pi\,\frac{R}{\varepsilon}\,\frac{w_s}{\varepsilon}\,\frac{W}{w_s}, \tag{B.1}$$

$$M_{3\,3}^s = M_{6\,6}^s = M_{9\,9}^s = M_{12\,12}^s = 2\,\pi\,\frac{R}{\varepsilon}\,\frac{w_s}{\varepsilon}\,\frac{W}{w_s}\left[\left(\frac{w_s}{\varepsilon}\right)^2\left(\frac{W}{w_s}\right)^2 + 4\left(\frac{R}{\varepsilon}\right)^2\right],$$

$$M_{1\,1}^r = M_{2\,2}^r = -\pi\left(\frac{R_r}{\varepsilon}\right)^2\frac{\varrho_r}{\varrho}, \qquad M_{3\,3}^r = -2\,\pi\,\frac{\varrho_r}{\varrho}\left(\frac{R_r}{\varepsilon}\right)^4$$

In order to express the 12-by-12 submatrix $\mathbf{K}_s^*$ (with components $K_{hk}^*$ and $h, k = 1, ..., 12$) it is necessary to introduce the dependent parameters $k_d/E_s$ and $k_\theta/(\varepsilon^2 E_s)$, which can be expressed as function the other parameters $\frac{k_d}{E_s} = f_d\left(\frac{R_r}{\varepsilon}\frac{\varepsilon}{R}, \frac{E_r}{E_s}, \nu_r\right)$ and $\frac{k_\theta}{\varepsilon^2 E_s} = f_\theta\left(\frac{R_r}{\varepsilon}\frac{\varepsilon}{R}, \frac{E_r}{E_s}, \nu_r\right)$ as reported in [35]. Then, the non null elements of the upper triangular part of $\mathbf{K}^s$ are expressed as follows:

$$K_{1\,1}^* = K_{2\,2}^* = K_{4\,4}^* = K_{5\,5}^* = K_{7\,7}^* = K_{8\,8}^* = K_{10\,10}^* = K_{11\,11}^* = 16\left(\frac{w_s}{\varepsilon}\right)^3 + 4\,\frac{w_s}{\varepsilon} \tag{B.2}$$

$$K_{3\,3}^* = K_{6\,6}^* = K_{9\,9}^* = K_{12\,12}^* = \frac{32}{3}\left(\frac{w_s}{\varepsilon}\right)^3 + 32\left(\frac{R}{\varepsilon}\right)^2\frac{w_s}{\varepsilon}$$

$$K_{1\,4}^* = K_{7\,10}^* = -2\,\frac{w_s}{\varepsilon}(\exp(-\iota k_1) + 1)$$

$$K_{1\,6}^* = K_{3\,4}^* = 4\,\frac{R}{\varepsilon}\frac{w_s}{\varepsilon}(\exp(-\iota k_1) - 1)$$

$$K_{1\,7}^* = K_{4\,10}^* = -8\left(\frac{w_s}{\varepsilon}\right)^3(\exp(-\iota k_2) + 1)$$

$$K_{1\,9}^* = K_{4\,12}^* = 4\left(\frac{w_s}{\varepsilon}\right)^3(\exp(-\iota k_2) - 1)$$

$$K_{2\,5}^* = K_{8\,11}^* = -8\left(\frac{w_s}{\varepsilon}\right)^3(\exp(-\iota k_1) + 1)$$

$$K_{2\,6}^* = K_{8\,12}^* = -4\left(\frac{w_s}{\varepsilon}\right)^3(\exp(-\iota k_1) - 1)$$

$$K_{2\,8}^* = K_{5\,11}^* = -2\,\frac{w_s}{\varepsilon}(\exp(-\iota k_2) + 1)$$

$$K_{2\,9}^* = K_{3\,8}^* = 4\,\frac{R}{\varepsilon}\frac{w_s}{\varepsilon}(\exp(-\iota k_2) - 1)$$

$$K_{3\,5}^* = K_{9\,11}^* = 4\left(\frac{w_s}{\varepsilon}\right)^3(\exp(-\iota k_1) - 1)$$

$$K_{3\,6}^* = K_{9\,12}^* = \frac{1}{3}\left[\left(\frac{w_s}{\varepsilon}\right)^2(1 + \exp(-\iota k_1)) - 6\left(\frac{R}{\varepsilon}\right)^2\left(1 + 4\,\frac{w_s}{\varepsilon}\exp(-\iota k_1)\right)\right]$$

$$K_{3\,7}^* = K_{6\,10}^* = -4\left(\frac{w_s}{\varepsilon}\right)^3(\exp(-\iota k_2) - 1)$$

$$K_{3\,9}^* = K_{6\,12}^* = \frac{1}{3}\left[\left(\frac{w_s}{\varepsilon}\right)^2(1 + \exp(-\iota k_2)) - 6\left(\frac{R}{\varepsilon}\right)^2\left(1 + 4\,\frac{w_s}{\varepsilon}\exp(-\iota k_2)\right)\right]$$

$$K^*_{5\,12} = K^*_{6\,11} = -4\,\frac{R}{\varepsilon}\frac{w_s}{\varepsilon}(\exp(-\iota k_2) - 1)$$

$$K^*_{7\,12} = K^*_{9\,10} = -4\,\frac{R}{\varepsilon}\frac{w_s}{\varepsilon}(\exp(-\iota k_1) - 1)$$

where $\iota$ denotes the imaginary unit. Finally, the non null components of the 3-by-3 $j$-th ($j = 13...16$) diagonal submatrices $\mathbf{K}^r_j$ (with components $K^r_{hk}$ and $h, k = 1, 2, 3$) read

$$K^r_{1\,1} = K^r_{2\,2} = -\frac{k_d}{E_s}, \qquad K^r_{3\,3} = -4\frac{k_\theta}{\varepsilon^2 E_s} \tag{B.3}$$

## Appendix C. Upper bound on the objective function

Each term $\Delta\omega_{hk,\mathcal{B}}$ or $\Delta\omega_{hk,\Gamma_p}$ can be written as the ratio

$$\frac{\mathcal{N}}{\mathcal{D}} = \frac{x_1 - x_2}{\frac{1}{2}(x_1 + x_2)} \tag{C.1}$$

where $x_1, x_2 \geq 0$ (for instance, $x_1 = \max_{\mathbf{k}\in\mathcal{B}} \omega_h(\boldsymbol{\mu}, \mathbf{k})$ and $x_2 = \min_{\mathbf{k}\in\mathcal{B}} \omega_k(\boldsymbol{\mu}, \mathbf{k})$ for the case of formula (10)). If, because of computational reasons, $\mathcal{B}$ (or $\Gamma_p$) is replaced by a grid, then the maximum and minimum over $\mathcal{B}$ (or $\Gamma_p$) are replaced by the maximum and minimum over its subset, possibly causing, respectively, a reduction $\epsilon_1$ of the maximum and an increase $\epsilon_2$ of the minimum. In other words, as a consequence of this discretization, $x_1$ and $x_2$ are replaced, respectively, by their approximations $\tilde{x}_1(\epsilon_1) = x_1 - \epsilon_1$ and $\tilde{x}_2(\epsilon_2) = x_2 + \epsilon_2$, with $\epsilon_1, \epsilon_2 \geq 0$ (*discretization errors*). Moreover, one gets also $\epsilon_1 \leq x_1$ because, in the context of the paper, the nondimensional frequencies are always non-negative. Hence, after the discretization, formula (C.1) is replaced by

$$\frac{\tilde{\mathcal{N}}(\epsilon_1, \epsilon_2)}{\tilde{\mathcal{D}}(\epsilon_1, \epsilon_2)} = \frac{\tilde{x}_1(\epsilon_1) - \tilde{x}_2(\epsilon_1)}{\frac{1}{2}(\tilde{x}_1(\epsilon_1) + \tilde{x}_2(\epsilon_1))} = \frac{x_1 - \epsilon_1 - x_2 - \epsilon_2}{\frac{1}{2}(x_1 - \epsilon_1 + x_2 + \epsilon_2)} \tag{C.2}$$

and one obtains

$$\frac{\tilde{\mathcal{N}}(\epsilon_1, \epsilon_2)}{\tilde{\mathcal{D}}(\epsilon_1, \epsilon_2)} \geq \frac{\tilde{\mathcal{N}}(0, 0)}{\tilde{\mathcal{D}}(0, 0)} = \frac{\mathcal{N}}{\mathcal{D}} \tag{C.3}$$

since the two first partial derivatives

$$\frac{\partial\left(\frac{\tilde{N}(\epsilon_1,\epsilon_2)}{\tilde{D}(\epsilon_1,\epsilon_2)}\right)}{\partial\epsilon_1} = \frac{-2(x_2 + \epsilon_2)}{\frac{1}{2}(x_1 - \epsilon_1 + x_2 + \epsilon_2)^2} \tag{C.4}$$

and

$$\frac{\partial\left(\frac{\tilde{N}(\epsilon_1,\epsilon_2)}{\tilde{D}(\epsilon_1,\epsilon_2)}\right)}{\partial\epsilon_2} = \frac{-2(x_1 - \epsilon_1)}{\frac{1}{2}(x_1 - \epsilon_1 + x_2 + \epsilon_2)^2} \tag{C.5}$$

are always non-positive. Finally, since the objective functions of the optimization problems considered in the paper are defined as the maxima among several terms of the form $\Delta\omega_{hk,\mathcal{B}}$ or $\Delta\omega_{hk,\Gamma_p}$, the bound (C.3) extends directly to the approximations of the objective functions. In other words, after the discretization, each objective function is replaced by an upper bound.